\documentclass[superscriptaddress, aps, pra, twocolumn, longbibliography]{revtex4-2}

\usepackage[utf8]{inputenc}

\usepackage{amsmath}
\usepackage{hyperref}
\usepackage{graphicx}
\usepackage{physics}
\usepackage{bbm}
\usepackage{url}
\usepackage{xcolor}
\hypersetup{
    colorlinks = true,
    linkcolor  = blue,
    citecolor  = blue,
    urlcolor   = blue,
}

\usepackage[normalem]{ulem}

\usepackage{lineno}

\graphicspath{{Figures/}}

\newcommand{\nbar}{\bar{n}}
\newcommand{\pr}{\text{Pr}}

\begin{document}
\title{Quantum illumination with multiplexed photodetection}
\author{Hao Yang}
\altaffiliation[email: ]{hao.yang@strath.ac.uk}
\affiliation{Department of Physics, University of Strathclyde, John Anderson Building, 107 Rottenrow, Glasgow G4 0NG, U.K.}
\author{Nigam Samantaray}
\affiliation{Department of Physics, University of Strathclyde, John Anderson Building, 107 Rottenrow, Glasgow G4 0NG, U.K.}
\affiliation{Quantum Engineering Technology Labs, H. H. Wills Physics Laboratory and Department of Electrical and Electronic Engineering, University of Bristol, BS8 1FD, UK}
\author{John Jeffers}
\affiliation{Department of Physics, University of Strathclyde, John Anderson Building, 107 Rottenrow, Glasgow G4 0NG, U.K.}
\date{\today}

\begin{abstract}
    The advantages of using quantum states of light for object detection are often highlighted in schemes that use simultaneous and optimal measurements. Here, we describe a theoretical but experimentally realizable quantum illumination scheme based on non-simultaneous and non-optimal measurements which can maintain this advantage. In particular, we examine the multi-click heralded two mode squeezed vacuum state as a probe signal in a quantum illumination process. The increase in conditioned signal intensity associated with multi-click heralding is greater than that from a single detector-heralded signal. Our results show, for lossy external conditions, the presence of the target object can be revealed earlier using multi-click measurements. We demonstrate this through sequential shot measurements based on Monte-Carlo simulation.
\end{abstract}

\maketitle

\section{Introduction}
\label{sec:introduction}
Quantum illumination provides advantages for object detection in a noisy environment: it relies upon quantum correlations between an entangled pair of beams to enhance the distinguishability of return signals from background noise alone \cite{Lloyd2008, Tan2008, Sanz2017}. In entangled state quantum illumination, radiation from one of the beams (signal) interrogates a target object whilst the other (idler) beam is stored locally until radiation reflected from any target returns. In an ideal scenario, the two correlated beams are measured simultaneously. If the correlations between the beams are nonclassical then the appropriate measurement strategy will provide a set of results whose statistics are forbidden in classical physics \cite{Bell1964}. Background radiation detected jointly with the stored local beam cannot provide such measurement statistics because it is uncorrelated with the idler, therefore by sending a quantum-correlated probe signal, the reflected signal can be identified from the background radiation as will be more distinguishable than sending an uncorrelated classical signal of equal intensity. The advantage can be inferred by calculating the error probability between two conditional states: the background state vs.~the object-reflected signal plus background. The minimum error probability achievable for an optimized measurement process is set by the Helstrom bound for two general states, but this bound is difficult to reach experimentally and difficult to compute for multiple observation trials of continuous-variable Gaussian states \cite{Helstrom1969, Audenaert2007, Calsamiglia2008}.

Simultaneous joint measurement of two beams is more difficult than detecting each beam independently -- in the latter case, no storage of light or consequent phase-lock are required. This lends itself to another measurement strategy in which idler detection occurs immediately at source, as a heralding detection, which provides a conditioning event upon the dispatched probe signal. This conditioning event affects the signal, causing changes in intensity or photon number from its pre-measurement state. Each individual unconditioned mode of the twin-beam contain thermal photon number statistics, but the interbeam quantum correlations mean that a count at an idler detector heralds an \emph{increase} in the mean photon number of the signal beam, which in turn increases the probability of a count at the signal detector \cite{Yang2020}. This model of quantum illumination under a non-simultaneous measurement strategy has applications in object-detection technology based on twin-beams of light produced through parametric down-conversion or by four-wave mixing, with one beam directly detected by a Geiger-mode (GM) avalanche photodiode (APD) in the form of a ``click" detection \cite{England2019}.

If counts are registered by the signal detector, then additional information from such heralding counts aids the search for signal photons reflected from the target, which are beneficial to target identification and subsequently ranging. Whilst the detection statistics obtained at the receiving signal detector cannot identify the presence of an object as well as a fully simultaneous measurement strategy of quantum illumination, they do provide advantages over sending a classical Poissonian or thermal signal beam of the same mean energy for target object identification. The advantages referred to above are even more marked if, instead of matching the energy of the quantum beam to the classical beam, the detection probability of the quantum and classical signal beams are matched. The differing statistics of the two allow the quantum signal to operate at higher mean photon number \cite{Yang:21}. Although the primary focus of our work is simply to emphasise the advantages of quantum illumination under simple detection schemes, the theory is potentially applicable to covert lidar operation which uses quantum states, one that can outperform a classical lidar at the same level of detectability by a third party.

\begin{figure}[!t]
    \centering
    \includegraphics[width=8.6cm]{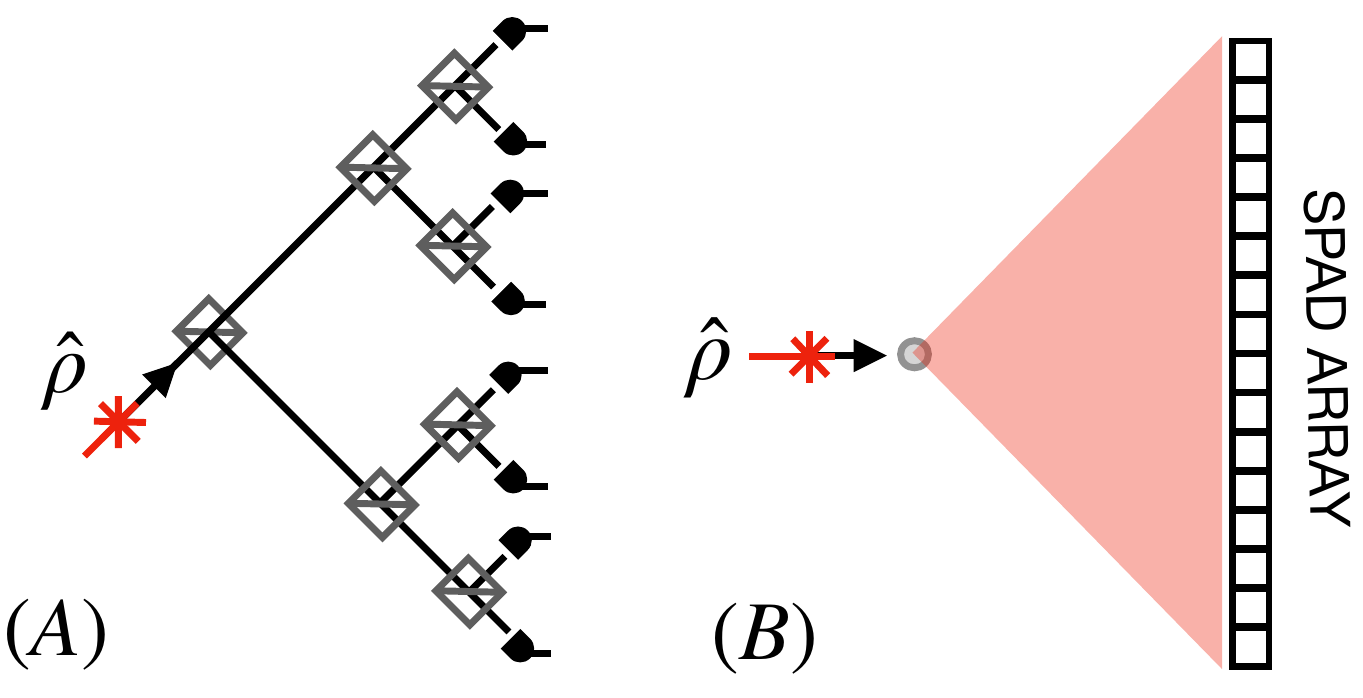}
    \caption{Two methods of multiplexed detection -- both are examples of spatial multiplexing. (A) Beamsplitter method: splitting a single mode $\hat{\rho}$, into $2N$ different paths by $2N-1$ beamsplitters via a symmetric arrangement. (B) Defocussing the beam evenly onto a SPAD array.}
    \label{fig:MultiplexMethods}
\end{figure}

The natural extension to the signal-boosting effect facilitated by single click detector heralding is to investigate the effect on the signal when multiple click detectors are used for heralding detection of the idler. Measurement of an entangled mode has been implemented as a method of state-engineering \cite{Paris2003}. A large number of click detectors will effectively act as a photon binning mechanism which would engineer photon number states \cite{Sperling2014a}, however we shall explore multiple click detection restricted to a few detectors. Multiple simultaneous click detection of a single mode can be achieved by arranging identical click detectors in a \emph{multiplex} configuration, either by beamsplitting a field mode onto multiple individual GM-detectors, or by defocussing onto a single photon avalanche photodiodes (SPAD) array (see Fig.~\ref{fig:MultiplexMethods}). As the click detector multiplex can be considered as a single unit, this detection scheme is applicable to receiver detection of the target object-reflected signal.

The paper is organised as follows. In Section \ref{sec:clickdet} we provide a summary of theoretical methods: starting with a brief review of click detection theory, before analysing the single-mode output state from multiplexed heralding click detection of an entangled twin-beam in Section \ref{sec:hwithcdet}. In Section \ref{sec:quantillum}, we calculate expressions for detection probabilities using the output multi-click heralded state for quantum illumination and highlight the advantages. The advantages can be further enhanced by matching the click probabilities with that of a coherent state, by tuning the mean photon number of the twin-beam, which we detail in \ref{sec:clickmatch}. We then follow with a numerical simulation of a sequentially updating detection strategy in Section \ref{sec:dettraj}, before discussion and some final concluding remarks.


\section{Click Detection Theory}
\label{sec:clickdet}
On-off or click detectors are useful, simple phase-insensitive measurement devices for light beams near the single photon level. Although the presence of a photon can be detected efficiently based on a click registered on these detectors, they are unable to distinguish between one or more photons with certainty. Another practical drawback is the dead-time, possibly lasting up to several nanoseconds, during which photons impinging in close succession to a successful detection will go unregistered. Through using arrays of fast SPADs \cite{Heilmann2016, Smith:16, Lundeen2009, Bodog:2020} in a multiplex, click detectors may achieve quasi-photon number resolution -- ``quasi" referring to the fact that true photon statistics can only be approximated by a finite number of click detectors.

\subsection{One Detector}
\label{subsec:onedet}
The single-mode click detector is a binary detector with quantum efficiency $\eta$ that fires (or does not) when met with radiation within its mode bandwidth. It has the following positive operator valued measure (POVM) element for its ``click" outcome
\begin{equation}
    \hat{\Pi}_{1,1}(\eta) = \mathbbm{1} - :e^{-\eta\hat{a}^\dagger\hat{a}}:,
\end{equation}
where the colons :: denotes normal-ordering of enclosed boson creation and annihilation operators, $\hat{a}^\dagger$ and $\hat{a}$. The expectation value of this operator for any detected field state with density matrix $\hat{\rho}$ provides the probability that the detector will fire. The term $:e^{-\eta\hat{a}^\dagger\hat{a}}:$ is the no-click POVM that is used to define the complementary click POVM. We may shed the normal-ordering and expand the exponential operator in the photon number basis via the following resolution \cite{Barnett2002}
\begin{equation}
\label{Eq:normfockexpansion}
    \hat{\Pi}_{1, 0}(\eta) = :e^{-\eta\hat{a}^\dagger\hat{a}}:\, = \sum_{n=0}^\infty (1-\eta)^n\ketbra{n},
\end{equation}
where $\ketbra{n}$ is the projection onto a state with $n$-photons. The above expression reveals that click detection is a phase-insensitive measurement and may only directly access the photon number distribution of a field state. Additional terms can be incorporated into the POVM if the detector has a significant intrinsic dark count probability. 

\subsection{Multiple Detectors}
\label{subsec:ngeqonedet}
Let us assume we branch a single-mode by splitting it onto $N$-multiple identical click detectors in a multiplexed scenario (see Fig.~\ref{fig:MultiplexMethods}). In such a setup, multiple simultaneous clicks may occur in a shot of the experiment. The combination of results for $k$ simultaneous clicks on $N$ detectors (thus $N - k$ detectors do not click) is provided by the binomial coefficient. Following this, the global multi-click POVM takes a binomial form
\begin{equation}
    \label{Eq:clickPOVM}
    \hat{\Pi}_{N,k}(\eta) = \binom{N}{k}:(e^{-\frac{\eta}{N}\hat{a}^\dagger\hat{a}})^{N-k}(\mathbbm{1}-e^{-\frac{\eta}{N}\hat{a}^\dagger\hat{a}})^{k}:,
\end{equation}
which reduces to single click detector POVM when $N=1$ \cite{Sperling2014a}. For a small number of detectors, this measurement would barely achieve photon number resolution, but in the limiting case where there is a large number of detectors, the click counting distribution
\begin{equation}
    \label{Eq:infClickPOVM}
    \lim_{N\rightarrow\infty}\left\langle\hat{\Pi}_{N, k}(\eta) \right\rangle=\, \left\langle:(\eta\hat{a}^\dagger\hat{a})^ke^{-\eta\hat{a}^\dagger\hat{a}}/k!:\right\rangle,
\end{equation}
becomes progressively more Poissonian for a large values of $N$ (Poisson limit theorem). Hence under perfect quantum efficiency, $k$-clicks given a very large number of detectors is equivalent to projection onto the Fock state $\ketbra{k}$ and will reproduce the true photon number distribution of the state \cite{Lee:2016, Miroslav:2019, Provaznik:20}. Another interpretation is that for such a large number of detectors, the intensity of the input light that falls on each one will be attenuated to such an extent that, at most, one photon would hit one detector. 

Again we may shed the normal-ordering of Eq.~\eqref{Eq:clickPOVM}, similar to  Eq.~\eqref{Eq:normfockexpansion} and expand the POVM in the photon number basis
\begin{multline}
    \label{Eq:clickPOVMfock}
    \hat{\Pi}_{N,k}(\eta) = \binom{N}{k}\\\times\sum_{n=0}^\infty \sum_{l=0}^k\binom{k}{l}(-1)^{k-l}\left[1-\eta\left(1-\frac{l}{N}\right)\right]^n\ketbra{n},
\end{multline}
with $k$ as the number of clicks at $N$-detectors and $l$ is an index over the the possible combinations of different sets of detectors firing to give $k$-clicks. If this is summed over all possible clicks from zero to $N$, the operators form a complete set
\begin{equation}
    \sum_{k = 0}^N \hat{\Pi}_{N,k}(\eta) = \mathbbm{1}.
\end{equation} 
We regain the one detector POVMs by setting $N = 1$
\begin{subequations}
\label{pi10}
    \begin{gather}
    \hat{\Pi}_{1,0}(\eta)=\sum_{n=0}^\infty (1-\eta)^n\ketbra{n},\\
    \hat{\Pi}_{1,1}(\eta) = \sum_{n=0}^\infty [1 - (1-\eta)^n]\ketbra{n},
    \end{gather}
\end{subequations}
hence for a general $N$-detector multiplex there are $N+1$ outcomes, all represented by POVMs. From here onwards a pair of numbered subscripts as in Eqs.~\eqref{pi10} above refer respectively to the number of detectors and the number of clicks.

\section{State Preparation With Multiple Detectors}
\label{sec:hwithcdet}
As multiplexed photodetection provides a larger set of measurement outcomes, measurement-based conditioning of an entangled state will provide an extended set of conditioned states. One mode of an entangled state is measured and, due to the stronger-than-classical correlations present in the entangled state, the unmeasured mode will be conditioned or ``engineered" into a different state. The entangled state we will consider in this report is the two mode squeezed vacuum (TMSV), which is a nonclassical state that can violate Bell's theorem under certain tests \cite{Banaszek1999, Kuzmich2000}, and is easy to produce experimentally \cite{Rarity:1991,Iskhakov:16,Brida:11,Krapick:2013, Whittaker:2017, Rarity:2017, Genovese:2009,Bambrilla:2004, Blancet:2008, Bambrilla:2008,Agafonov:2010,Bondani:2007,Perina:2012} as it is a generalized output of nondegenerate spontaneous parametric down-conversion (SPDC). 

The TMSV state is (ignoring a possible phase attached to each term)
\begin{equation}
    \label{Eq:TMSV}
    \ket{\Psi}_{I, S} = \frac{1}{\sqrt{1+\nbar}}\sum_{n=0}^\infty \left(\frac{\nbar}{1+\nbar}\right)^{n/2} \ket{n, n},
\end{equation}
where $\ket{n,n}$ is the state containing $n$ photons in each of the two modes denoted by subscripts: $I$ -- idler and $S$ -- signal. Each of the modes individually has mean photon number 
\begin{equation}
    \nbar = \sinh^2r,
\end{equation} with $r$ as the squeezing amplitude which depends on the nonlinearity of the crystal and the intensity of the pump light. If one of the beams is measured independently, without information regarding the other, then the measurement outcomes will replicate Bose-Einstein statistics of a thermal state with mean photon number $\nbar$ (such a thermal state will be referred to in this report as the ``unconditioned" state).

By using a single click detector in the idler to condition the TMSV signal mode, the remaining signal mode will be engineered into a vacuum suppressed state if a successful click occurs. The vacuum suppressed state has the following density operator
\begin{multline}
    \label{Eq:oneclickstate}
    \hat{\rho}_{1,1} = \frac{1}{\pr_{1, 1}} \tr_I\left(\hat{\Pi}_{1,1}\otimes\mathbbm{1} \ketbra{\Psi}_{I,S}\right)\\
     = \mathcal{N}_{1,1}\sum_{n = 0}^\infty \big[1-(1-\eta)^n\big]\left(\frac{\nbar}{1+\nbar}\right)^n\ketbra{n},
\end{multline}
which is a nonclassical state that contains no vacuum probability, with $\mathcal{N}_{1,1} = [(1+\nbar)\pr_{1,1}]^{-1}$ as the normalization factor and $\Pr_{1,1} = 1 - (1 + \nbar\eta)^{-1}$ as the heralding probability for a single click detector. This state has a conditioned mean photon number of 
\begin{equation}
    \nbar_{1,1} = \tr(\hat{a}^\dagger \hat{a}\hat{\rho}_{1,1}) =  \nbar + \frac{1+\nbar}{1+\eta\nbar},
\end{equation}
evidently an increase from its unconditioned number. In the limit of unit quantum efficiency the mean is increased by one, regardless of the unconditioned state mean photon number. This can be a large fractional increase for low $\bar{n}$ and is the basic effect exploited for optical quantum illumination.

\begin{figure}[!t]
    \includegraphics[width = 8.6cm]{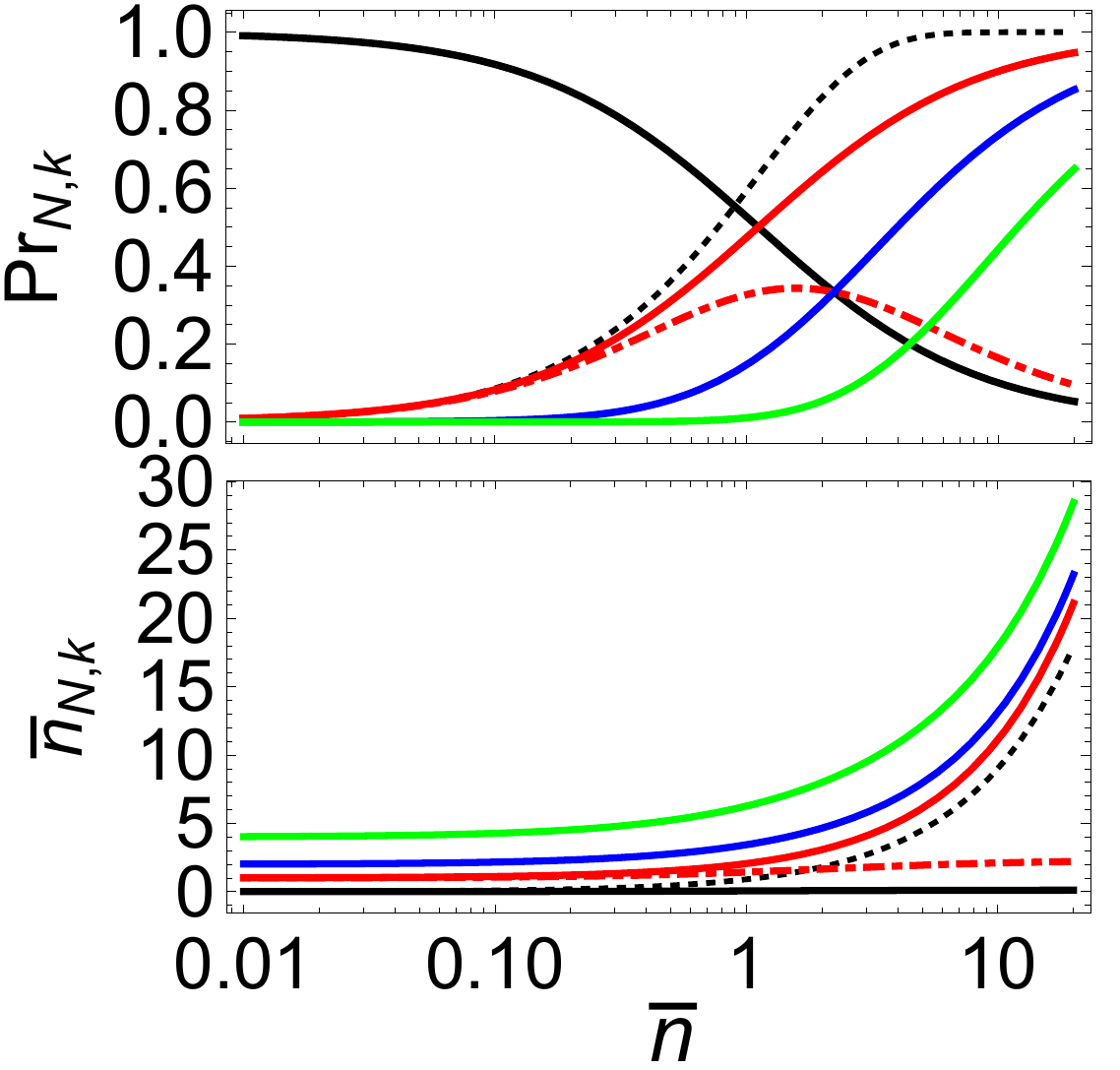}
    \caption{(Color online) Top: Multiplex heralding click probability of TMSV vs.~mean photon number of the unconditioned state, $\bar{n}$, with detection efficiency of $\eta=0.95$. Black dotted line corresponds to the single detector click probability for a coherent state $\ket{\alpha}$. Black solid line is the no-click probability for the thermal state. Red dash-dot line is the heralding click probability $\Pr_{2, 1}$. The rest of the solid lines are click probabilities for $N = k$ outcomes: green (light gray) for $\Pr_{4,4}$; blue (dark gray) for $\Pr_{2,2}$ and red (gray) for $\Pr_{1,1}$. The intersection points show equally probable heralding click probabilities for differing number of $N$ and $k$ given $\nbar$, e.g.~$\pr_{4,4} = \pr_{2,1}$ for $\nbar = 5.4$. Bottom: Mean number of photons in the conditioned signal mode $\hat{\rho}_{N,k}$ after a successful multiplexed click heralding detection, as a function of unconditioned state mean photon number. Black solid line and red dash-dot line are for states $\hat{\rho}_{1,0}$ and $\hat{\rho}_{2,1}$ respectively. The rest of the solid lines for $N = k$ from bottom to top follow reverse order with respect to top plot.}
    \label{fig:HeraldandAvg}
\end{figure}

Building on the results of a single click detector, the aim now is to obtain, through multiplexed photodetection, a wider array of nonclassical states with increased mean photon numbers that cannot be achieved via a single click detector. By changing the POVM in Eq.~\eqref{Eq:oneclickstate} to $\hat{\Pi}_{N,k}$ (hence the heralding probability to $\Pr_{N,k}$), the multi-click measurement of the TMSV idler mode will herald the general output state with the following density matrix
\begin{multline}
\label{Eq:outputstate}
    \hat{\rho}_{N, k} = \frac{1}{\Pr_{N,k}} \tr_I\bigg(\hat{\Pi}_{N,k} \otimes \mathbbm{1} \ketbra{\Psi}_{I,S}\bigg) \\ 
    = \mathcal{N}_{N,k}\sum_{l=0}^k \binom{k}{l} (-1)^{k-l}(1+\bar{m}_{N,l})\hat{\varrho}\left[\bar{m}_{N,l}\right],
\end{multline}
with 
\begin{equation}
\hat{\varrho}\left[\bar{m}_{N,l}\right] = \frac{1}{1+\bar{m}_{N, l}}\sum_{n = 0}^\infty\left(\frac{\bar{m}_{N, l}}{1 + \bar{m}_{N, l}}\right)^n \ketbra{n},
\end{equation}
as the thermal state density matrix with a scaled mean photon number
\begin{equation}
\label{Eq:scaledcoeff}
    	\bar{m}_{N,l} = \frac{\bar{n}-\bar{n}\eta\left(1-\frac{l}{N}\right)}{1+\bar{n}\eta\left(1-\frac{l}{N}\right)}.
\end{equation}
Here the subscript $l$ is used instead of $k$ because the thermal state is inside the $k$ sum in Eq.~\eqref{Eq:outputstate} (note: we henceforth use the notation $\varrho[\nbar]$ to denote the thermal state with mean photon number $\nbar$).
The normalization
\begin{equation}
    \mathcal{N}_{N,k} = \left[ \sum_{l=0}^k\binom{k}{l}(-1)^{k-l}(1+\bar{m}_{N,l})\right]^{-1},
\end{equation}
is related to the heralding click probability via
\begin{equation}
    \label{Eq:heraldprob}
    \pr_{N, k} = \binom{N}{k}\frac{1}{\mathcal{N}_{N,k}(1+\nbar)}.
\end{equation}
As previously mentioned in the introduction, a single mode of the TMSV appears thermal when it is observed independently of the other mode, hence the photon number distribution of the unconditioned mode has a Bose-Einstein distribution: for which the highest probability occurs for the vacuum state, being $(1+\nbar)^{-1}$. The heralding multiplex measures the idler mode independently of the signal mode, hence the multiplex heralding click probability $\Pr_{N,k}$ is that of a thermal state. Multi-clicks are therefore unlikely to happen when measuring TMSV with a low mean photon number, but become more likely with increasing mean photon number as shown by Fig.~\ref{fig:HeraldandAvg}.

\begin{figure}[!t]
    \centering
    \includegraphics[width=8.6cm]{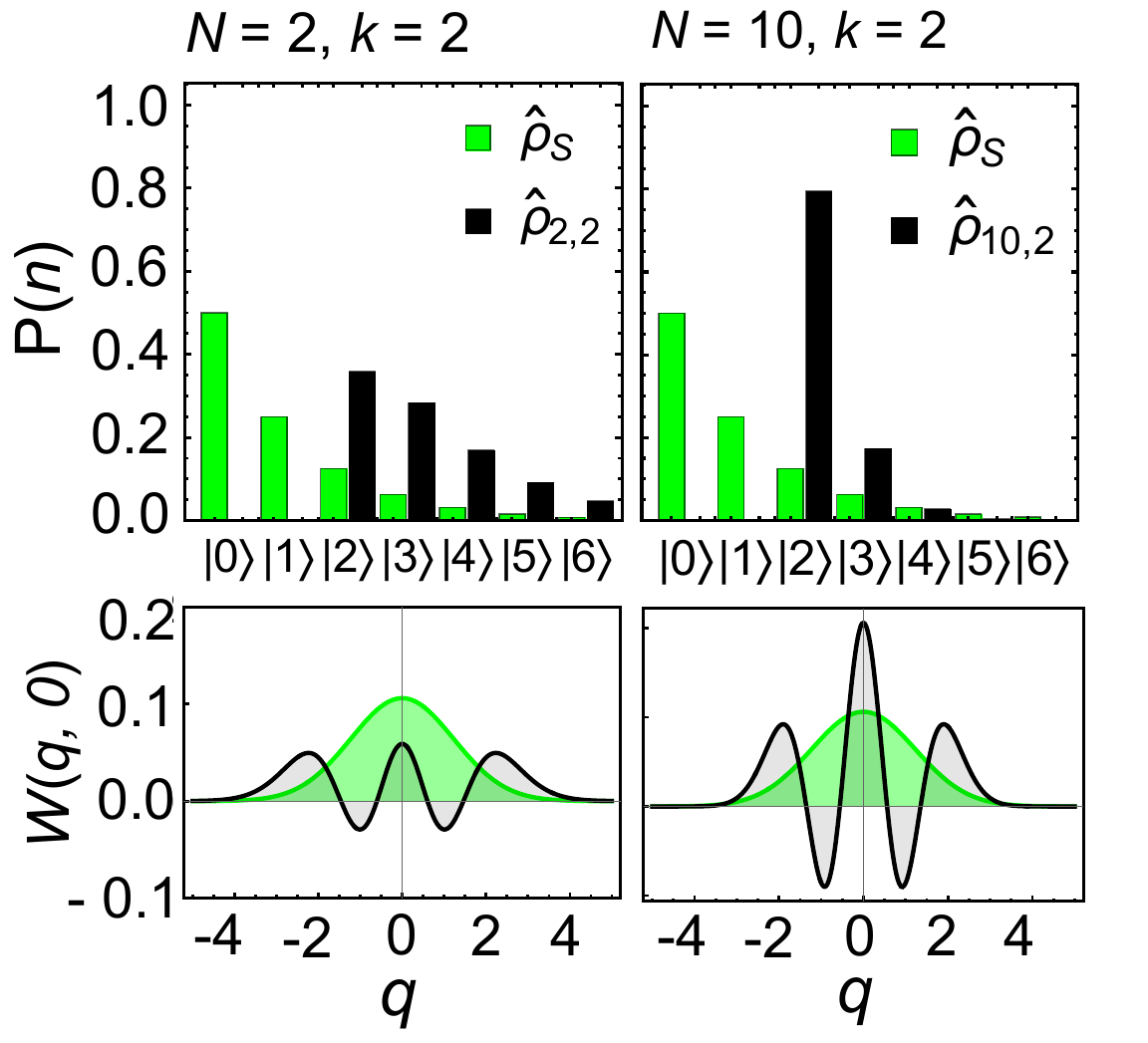}
    \caption{Top row: photon number distributions of the multiplex click heralded TMSV $\hat{\rho}_{N, k}$, heralded by two-clicks at the idler, compared with that of the unconditioned signal mode $\hat{\rho}_S$ with $\bar{n}=1$. Bottom row: phase space distribution (Wigner function) slices $W(q,0)$ through phase space $(q,p)$ of the corresponding states showing the transformation of a thermal state into a nonclassical one. Left column: $N = 2$ heralding detectors. Right column: $N = 10$ heralding detectors.}
    \label{fig:IncreasingN}
\end{figure}

The photon statistics of the multiplexed click heralded TMSV shown by Eq.~\eqref{Eq:outputstate} are those of a weighted binomial sum of $k$ thermal states, $\hat{\varrho}$, with a scaled mean photon number $\bar{m}_{N,l}$, that is a function of the detector efficiency and the number of detectors. Some features of such a superposition of thermal states are shown in Fig.~\ref{fig:IncreasingN}, which shows the normalized photon number distribution and a corresponding phase space distribution (Wigner function) slice. The photon number distributions show that successful multiplexed heralding of the TMSV produces a state with vanishing photon number probabilities up to one fewer than the number of heralding clicks, reflecting the fact that the state has a higher mean photon number than its unheralded counterpart. The state tends to a Fock state \cite{Sperling2014a} if the number of detectors is much greater than the number of heralding clicks, an effect that can be seen by comparing the two top panels in Fig.~\ref{fig:IncreasingN}. The nonclassicality of the state is often demonstrated by the negative values of the Wigner function \cite{Barnett2002, Schleich2005}, which is a quasi-probability distribution that represents a density matrix in position-momentum space. The bottom figures in Fig.~\ref{fig:IncreasingN} show the Wigner function of $\hat{\rho}_{N, k}$ through the $p = 0$ plane, because this particular Wigner function has rotational symmetry about the $z$-axis. The nonclassicality of the Wigner function, however, is a sufficient (but not necessary) condition of nonclassicality \cite{HILLERY:1984, Miroslav:2021, Kenfack2004, William:2008}. 

\begin{figure}[!t]
    \includegraphics[width = 8.6cm]{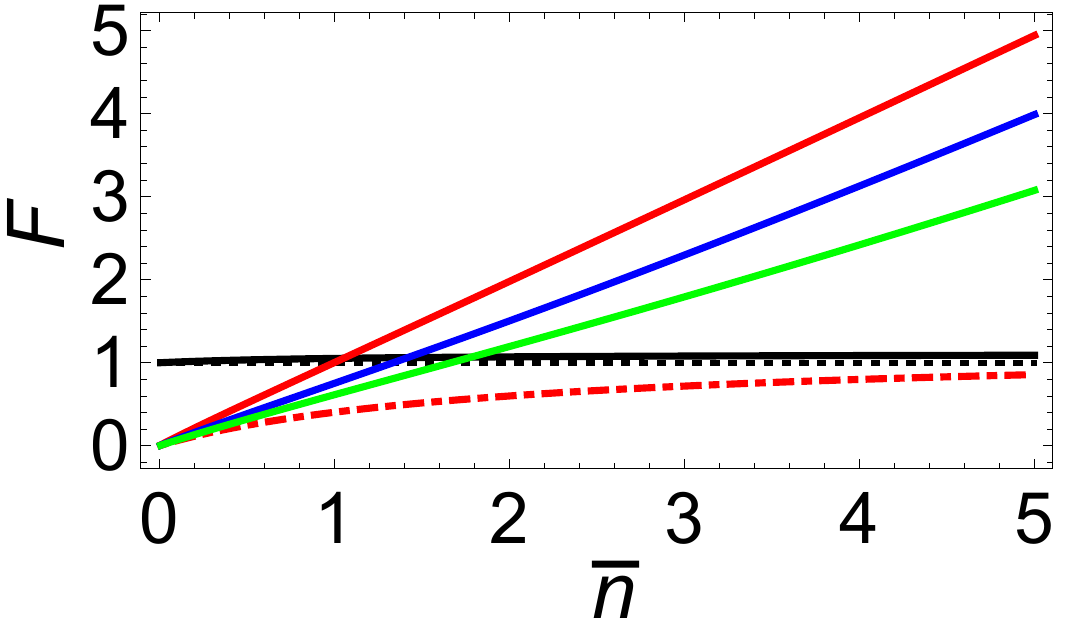}
    \caption{(Color online) Fano factor of the output state $\hat{\rho}_{N,k}$ vs.~mean photon number of the unconditioned state, $\bar{n}$, with detection efficiency $\eta=0.9$. Black dotted line corresponds to coherent state Fano factor that demarcates the domain between sub-Poissonian vs.~super-Poissonian statistics. Black solid line show the Fano factor for state $\hat{\rho}_{1, 0}$, and red dash-dot line for $\hat{\rho}_{2, 1}$. The rest of the solid lines from bottom to top are for the following states: green (light gray) for $\hat{\rho}_{4, 4}$; blue (dark gray) for $\hat{\rho}_{2, 2}$ and red (gray) for $\hat{\rho}_{1, 1}$.}
    \label{fig:FanoF}
\end{figure}

Another measure used to quantify nonclassicality in the photon number distribution is via the Fano Factor
\begin{equation}
    F=\frac{\langle \hat{n}^{2}\rangle-\langle \hat{n}\rangle^{2}}{\langle \hat{n}\rangle},	
\end{equation}	
where $\hat{n} = \hat{a}^\dagger\hat{a}$ is photon number operator; the Fano factor is evidently the variance divided by the expectation value of this operator. $F = 1$ indicates Poissonian statistics of coherent state regardless of detection loss $\eta$, whereas states with $0 \leq F < 1$ and $F > 1$ show sub-Poissonian (nonclassical) and super-Poissonian (thermal-like) statistics respectively. In the special case where all heralding detectors fire ($k = N$), the photon number variance of the heralded TMSV is sub-Poissonian for low $\bar{n}$ until it becomes super-Poissonian for higher $\bar{n}$. The Fano factor then show a corresponding transition from nonclassical to thermal as $\bar{n}$ varies from low to high values as evident from Fig.~\ref{fig:FanoF}. More detectors require higher $\nbar$ for the output state to crossover into the super-Poissonian domain, but the fractional increase in nonclassicality is smaller for each increase in $k$ and $N$.

Perhaps the most useful physical effect that multiplex click heralding of TMSV facilitates is the increase in mean photon number in the signal mode from its unconditioned form, with selected examples shown by Fig.~\ref{fig:HeraldandAvg}. This will be the main effect exploited for quantum illumination. The effect is a direct consequence of the photon number correlation between signal and idler beams of the TMSV: $k$-heralding clicks occurring at the idler mode will force the signal mode from a thermal state into a (nonclassical) state with boosted mean photon number. This relative increase becomes larger if $\nbar$ is small, shown in the bottom panel of Fig.~\ref{fig:HeraldandAvg} where there is a ``jump" caused by the heralding clicks in the limit of very low $\nbar$. When all of the detectors click ($k = N$), the increase in the signal mean photon number is maximal for an $N$-detector multiplex (notice that in Fig.~\ref{fig:HeraldandAvg}, $\nbar_{1,1} > \nbar_{2,1}$: if single click results are considered, then single clicks are more likely to occur if the detector is not in a multiplex). Although heralding can bring an increase in mean photon number in classically correlated states also \cite{Zhai:13}, it is not possible to obtain the large fractional boosts obtainable via click heralding the TMSV.

Strongly correlated temporal modes are prerequisite for efficient heralding click detection and in the generation of a conditioned signal state. Usually in experiments involving state generation from twin beams, such modes are produced and manipulated using picosecond pulse width lasers at repetition rates of nanoseconds. The conditioned states are then analysed on a pulse to pulse basis, triggered by multiple clicks at heralding detectors, in order to discern signal photons from other ambient stray light and noise sources. This is facilitated by time gating in the heralding click detection, which is realized by programmable electronics and optical switching at nanosecond speeds \cite{Brida:11, Rarity:2017}. The repetition rate of the laser is synchronised to that of the optical switch so that the time correlation between heralding clicks and the conditioned signal state can be ensured. Dark counts, which are thermal in origin, introduce errors in measurements; the smaller their rate, the better the accuracy of the measurement. For ultra-cooled detectors and pulsed operation in the nanosecond measurement time window, the dark counts can be as low as $10^{-8}$ mean counts per single detector \cite{Rarity:2000}, but they add with the number of multiplexed detectors. For reasonable numbers of detectors, the number of dark counts is still very low and can be neglected in experiments. Henceforth, unless explicitly mentioned, dark counts are ignored in the analysis.


\section{Quantum Illumination with Multiplexed Photodetection}
\label{sec:quantillum}
In the previous sections we have characterized the multiplex click heralded TMSV states and provided POVM expressions for the click detector multiplex -- now we shall apply them to our analysis of the quantum illumination model with multiplexed photodetection.

\subsection{Quantum Illumination Model}
\begin{figure}[!t]
	\centering
	\includegraphics[width=8.6cm]{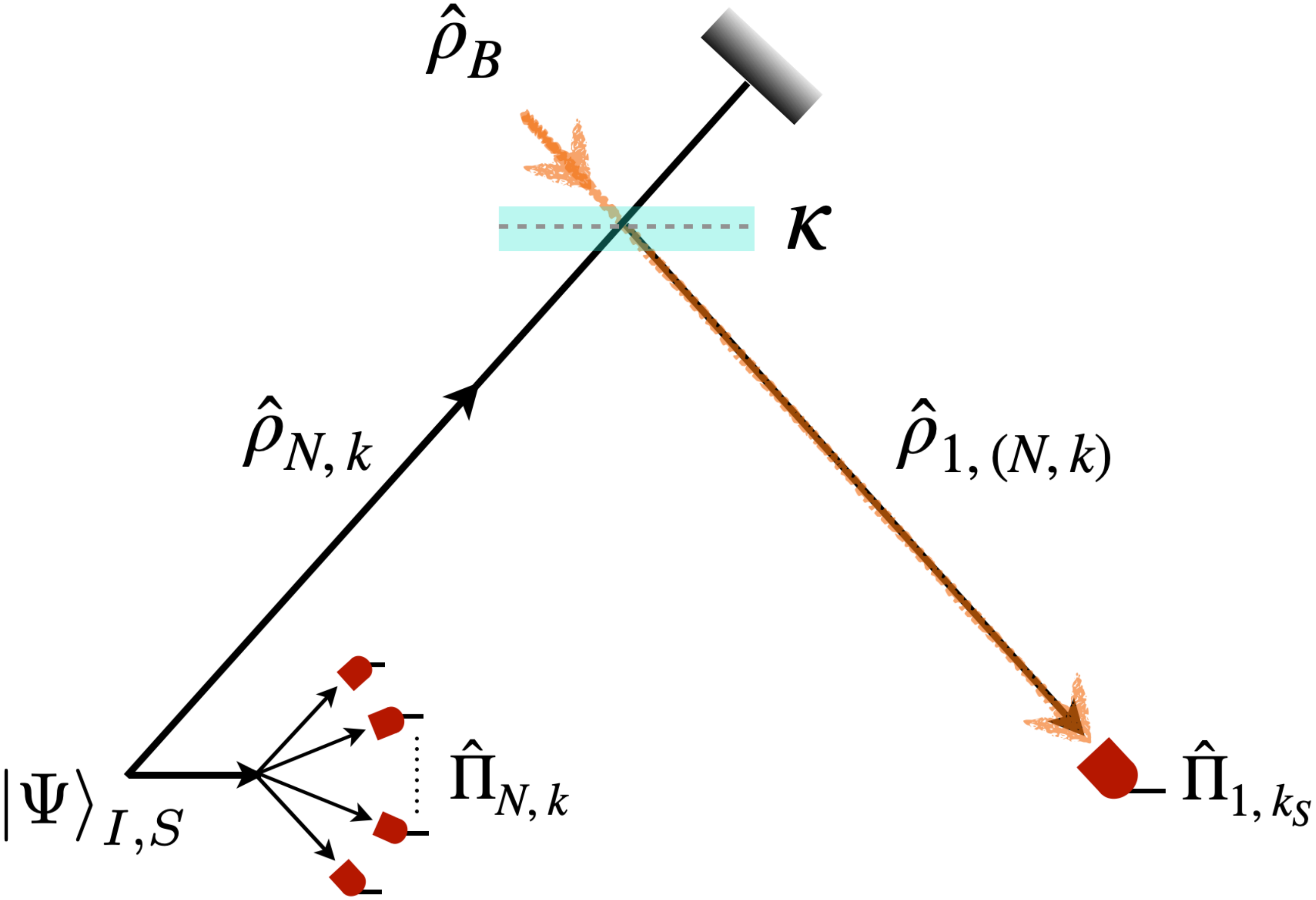}
	\caption{Schematic of quantum illumination with multiplexed photodetection showing scenario $H_1$ where the target is present: heralded state $\hat{\rho}_{N, k}$ is generated from TMSV $\ket{\Psi}_{I,S}$ by multiplexed click detection of its idler mode. The target object is modelled by a beamsplitter of reflectivity $\kappa$, which mixes the heralded signal with background thermal mode $\hat{\rho}_B$. The reflected state is denoted $\hat{\rho}_{1, (N, k)}$ and is measured by a single click detector, although in the general case this detector may be replaced with another detector multiplex. The target absent case $H_0$ would be equivalent to removing the beamsplitter in the diagram or by setting $\kappa = 0$: in this case only the background mode will be measured.}
	\label{fig:Schematic}
\end{figure}

If we send a single mode signal state $\hat{\rho}_{S}$ to try and sense the presence of a target object with reflectivity $\kappa$, embedded in a thermal background field, then based on the presence or absence of the target object two possible conditional states of the field must be considered for the receiving measurement. These are $\hat{\rho}_0$, which is the conditional state when the target is absent, and $\hat{\rho}_1$, which is the state when the target is present. They are, respectively, representations of the field under the null hypothesis ($H_0$) and the alternate hypothesis ($H_1$). The density operators of these two conditional states are
\begin{subequations}
	\begin{eqnarray}
	&\hat{\rho}_0 = \hat{\varrho}[\nbar_B], \label{seq:rho0}\\
	&\hat{\rho}_1 = \tr_E \left( \hat{U}_{\kappa}\hat{\rho}_{S}\otimes\hat{\varrho}[\nbar_B/(1-\kappa)]\hat{U}_{\kappa}^\dagger\right),
	\label{seq:rho1}
	\end{eqnarray}
\end{subequations}
where $\hat{\varrho}[\bar{n}_{B}]$ is a thermal state density matrix with mean photon number $\bar{n}_{B}$, modelling background radiation. In the $\hat{\rho}_1$ expression, the state contains signal and background photons, because the signal mode has been mixed with the thermal background mode with a re-scaled mean photon number $\bar{n}_{B}/(1-\kappa)$, in order to keep the received background noise independent of the object reflectivity. Reflection is modelled as coupling of the signal and background modes via a mixing process that is facilitated by the two-mode rotation (beamsplitter) unitary operator
\begin{equation}
    \hat{U}_{\kappa}= e^{i\theta(\hat{a}_S^\dagger\hat{a}_B + \hat{a}_S\hat{a}_B^\dagger)},
\end{equation}
with $\hat{a}_S$ and $\hat{a}_B$ are the field operators of the signal and background modes. The reflectivity is a function of the mixing angle
\begin{equation}
    \kappa = \cos^2\theta,
\end{equation}
hence $0 < \kappa < 1$. As the beamsplitter unitary couples two input modes to produce two outputs, the environment mode containing the unreflected signal will not be detected, therefore in the expression $\hat{\rho}_1$ it is traced out via $\tr_E$. A schematic of the quantum illumination model is shown in Fig.~\ref{fig:Schematic}.

The quantum probe signal we shall use is the multi-click heralded TMSV $\hat{\rho}_{N,k}$ described previously by Eq.~\eqref{Eq:outputstate}. For all single mode probe signals, $\hat{\rho}_0$ retains the same expression because it will be the state measured given signal loss. By substituting Eq.~\eqref{Eq:outputstate} into Eq.~\eqref{seq:rho1}, we have
\begin{equation}
    \hat{\rho}_{1, (N,k)} = \tr_E \left( \hat{U}_{\kappa}\hat{\rho}_{N,k}\otimes\hat{\varrho}[\nbar_B/(1-\kappa)]\hat{U}_{\kappa}^\dagger\right),
\end{equation}
with indices $1, (N, k)$ denoting solely the state for $H_1$, which contains the multi-click heralded TMSV signal, $\hat{\rho}_{N,k}$. As this is a weighted sum of thermal states, the expression for $\hat{\rho}_{1, (N,k)}$ will be a weighted sum of transformed thermal states. The transformation for $\hat{\varrho}[\bar{m}_{N,l}]$ alone gives
\begin{equation}
    \tr_E\left( \hat{U}_{\kappa}\hat{\varrho}[\bar{m}]\otimes\hat{\varrho}[\nbar_B/(1-\kappa)]\hat{U}_{\kappa}^\dagger\right) = \hat{\varrho}[\kappa\bar{m}+\nbar_B],
\end{equation}
which in effect modifies the input thermal state by reducing the mean by the reflection factor $\kappa$ and adding the transmitted background thermal mean $\bar{n}_{B}$. Hence the full expression for $\hat{\rho}_{1, (N,k)}$ is
\begin{multline}
    \label{Eq:rho1rhoNk}
    \hat{\rho}_{1, (N,k)} = \mathcal{N}_{N, k} \\
    \times \sum_{l=0}^k \binom{k}{l} (-1)^{k-l}(1+\bar{m}_{N,l}) \hat{\varrho}\left[\kappa\bar{m}_{N,l}+\nbar_B\right].
\end{multline}
Usually, the next steps for proof of quantum illumination advantage over classical illumination would be to calculate absolute error bounds (as a function of a chosen parameter, e.g.~mean photon number of signal) using such pairs of conditional states $\hat{\rho}_0$ and $\hat{\rho}_{1, (N, k)}$, and compare with such bounds produced from a coherent state signal. The absolute error bounds are directly related to the discrimination error achievable from \emph{optimal} measurements. The click detector is obviously not an optimal measurement, but as it clicks with different probabilities for $\hat{\rho}_0$ and $\hat{\rho}_1$ (given that they are not the same state), we can use click detection for measurement of the conditional states as a practical implementation of hypothesis testing. 

\subsection{Signal Detector Probabilities}
We will first model a second click detector multiplex as the receiving signal detector in order to obtain a general expression that includes both heralding and signal detector variables. The overall strategy is to perform measurements and infer from click counts the probability of target presence (or absence). In order to do so, we define the conditional probability of obtaining $k_S$ clicks at an $N_S$-multiplex signal detector as (subscript $S$ denotes signal detector variables)
\begin{equation}
    \label{Eq:conditonalprob}
    \pr_{N_S}\left(k_S|H_i\right) = \tr (\hat{\Pi}_{N_S, k_S}\hat{\rho}_i),
\end{equation}
given $i = 0, 1$ for either of the conditional states, where $\hat{\Pi}_{N_S, k_S}$ is the POVM of the detector of quantum efficiency $\eta_S$, given by Eq.~\eqref{Eq:clickPOVMfock}. The above expression is true for all numbers of idler clicks $k$ so we omit this variable here unless specifically required. From this probability expression, explicitly shown as a conditional probability, we can calculate the posterior probability of a state, given a multi-click outcome, via Bayes' Law 
\begin{equation}
    \label{Eq:bayeslaw}
    \pr_{N_S}\left(H_i| k_S\right) = \frac{\Pr(H_i)\pr_{N_S}\left(k_S|H_i\right)}{\pr_{N_S}(k_S)},
\end{equation}
where $i = 0, 1$. The normalization $\pr_{N_S}(k_S)$ is the weighted sum of all conditional probabilities under click outcome $\hat{\Pi}_{N_S, k_S}$
\begin{equation}
    \pr_{N_S}\left(k_S\right) = \sum_{i = 0}^1\pr(H_i) \pr_{N_S}\left(k_S|H_i\right),
\end{equation}
where $\pr(H_i)$ is the prior probability the estimation of which is a non-trivial problem. In Sec.~\ref{sec:dettraj} we describe a method for estimation of this value in a sequential measurement process. 

The multi-click probabilities at a signal-receiving detector are found from the two hypothesis states,
\begin{multline}
    \label{Eq:h0click}
    \pr_{N_S}( k_S|H_0) = \binom{N_S}{k_S}\frac{1}{1+\nbar_B}\\ \times \sum_{l=0}^k \frac{\binom{k_S}{l_S}(-1)^{k_S-l_S}}{1-\frac{\nbar_B}{1+\nbar_B}\left(1-\eta_S(1-l_S/N_S)\right)},
\end{multline}
\begin{multline}
    \label{Eq:h1quantclick}
    \pr_{N_S}( k_S|H_1) = \mathcal{N}_{N,k}\binom{N_S}{k_S}\\ \times\sum_{l=0}^{k}\sum_{l_S=0}^{k_S}\frac{\binom{k}{l}\binom{k_S}{l_S}(-1)^{k+k_S-l-l_S}\left(\frac{1+\bar{m}_{N,l}}{1+\kappa\bar{m}_{N,l}+\nbar_B}\right)}{1-\left(\frac{\kappa\bar{m}_{N,l}+\nbar_B}{1+\kappa\bar{m}_{N,l}+\nbar_B}\right)(1-\eta_S(1-l_S/N_S))},
\end{multline}
which are the expectation values of the multi-click POVM with states $\hat{\rho}_0$ and $\hat{\rho}_{1, (N, k)}$ respectively. The latter equation is implicitly a coincidence probability $\Pr_{N, N_S}(k, k_S|H_1)$, however we have chosen to define the above probabilities in terms of one detector variable at a time as $\Pr_{N_S}(k_S|H_1)$, therefore the idler detection variables are encoded in the density matrix $\hat{\rho}_{1,(N,k)}$ instead. For low-reflectivity targets, a very small fraction of the probe photons will reach the detector and the return state predominantly contains thermal background photons for which additional multi-clicks events given state $\hat{\rho}_{1, (N, k)}$ will be rare, but not quite as rare as if only single click heralding had been used on the idler beam. Therefore, it will be sufficient to analyze click probabilities of a single receiving detector, which can be easily found from Eq.~\eqref{Eq:h0click} and Eq.~\eqref{Eq:h1quantclick} by setting $N_S = 1$. If the target is absent the state $\hat{\rho}_0$ reaching the detector is merely the thermal background: the probability of no-click is 
\begin{equation}
    \pr_1(0|H_0) = \frac{1}{1 +\eta_S\nbar_B},
\end{equation}
and in this situation therefore, the single click (sometimes called false-alarm) probability is
\begin{equation}
	\pr_1(1|H_0) = \frac{\eta_S\nbar_B}{1 + \eta_S\nbar_B}.
\end{equation}
Neither is a function of target reflectivity, only of receiver quantum efficiency. Hence they hold for all single-mode probe states. The no-click probability for $\hat{\rho}_1$ with multiplex click heralded state at a single receiving click detector is 
\begin{multline}
    \pr_1(0 | H_1) = \mathcal{N}_{N,k} \\ \times \sum_{l = 0}^k \frac{(-1)^{k-l}\left(\frac{1+\bar{m}_{N,l}}{1 + \kappa\bar{m}_{N, l} + \nbar_B}\right)}{1 - \left(\frac{\kappa\bar{m}_{N,l}+\nbar_B}{1 + \kappa\bar{m}_{N, l} + \nbar_B}\right)(1-\eta_S)},
\end{multline}
obtained via Eq.~\eqref{Eq:h1quantclick}; the signal detector click probability is again $\Pr_1(1 | H_1) = 1 - \Pr_1(0 | H_0)$.

\begin{figure}[!tb]
	\centering
	\includegraphics[width=8.6cm]{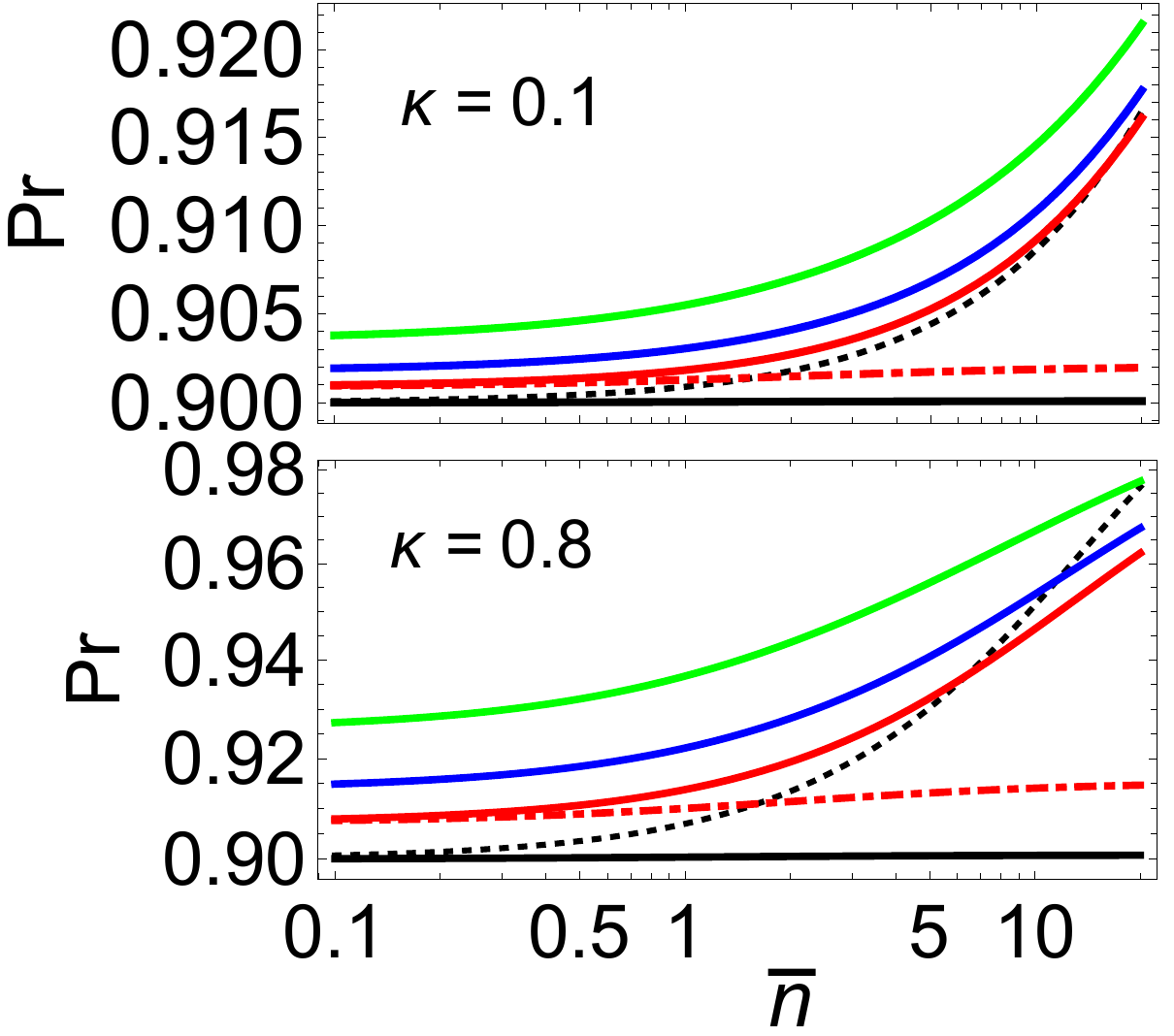}
	\caption{(Color online) Click probability of a single receiving signal detector measuring returning state $\hat{\rho}_{1, (N,k)}$ with detector efficiencies $\eta_S = \eta = 0.9$, against mean photon number of the unconditioned state $\nbar$ and background noise $\nbar_B = 10$. Black dotted line is the performance of coherent state. In both plots we show click probabilities given object presence $\Pr_1(1|H_1)$, for the following heralded signal states: black solid line for $\hat{\rho}_{1,(1,0)}$, red dash-dot for $\hat{\rho}_{1, (2,1)}$, red solid line (gray) for $\hat{\rho}_{1,(1,1)}$, blue solid line(dark gray) for $\hat{\rho}_{1,(2,2)}$ and green solid line (light gray) for $\hat{\rho}_{1,(4,4)}$.}
	\label{fig:MultiplexedDetectors}
\end{figure}

The probability of obtaining a single click, given target object presence, as a function of signal mean photon number is shown in Fig.~\ref{fig:MultiplexedDetectors}. For a lossy scenario of low reflectivity of $\kappa = 0.1$ and background mean photon number $\nbar_B = 10$, we witness click probability enhancement due to the multi-click heralded states exceeding that produced by a coherent state signal. Successive number of simultaneous clicks further enhance signal detector click probability, as increases to the mean photon number in the probe state are produced; maximum enhancement is found when all detectors in the heralding multiplex click ($N = k$). Furthermore, it is evident that for low $\nbar$, using multi-click heralded states for quantum illumination is more beneficial as the relative increase in the mean number of signal photons is greater, as shown by Fig.~\ref{fig:HeraldandAvg}. 

\begin{figure}[!tb]
	\centering
	\includegraphics[width=8.6cm]{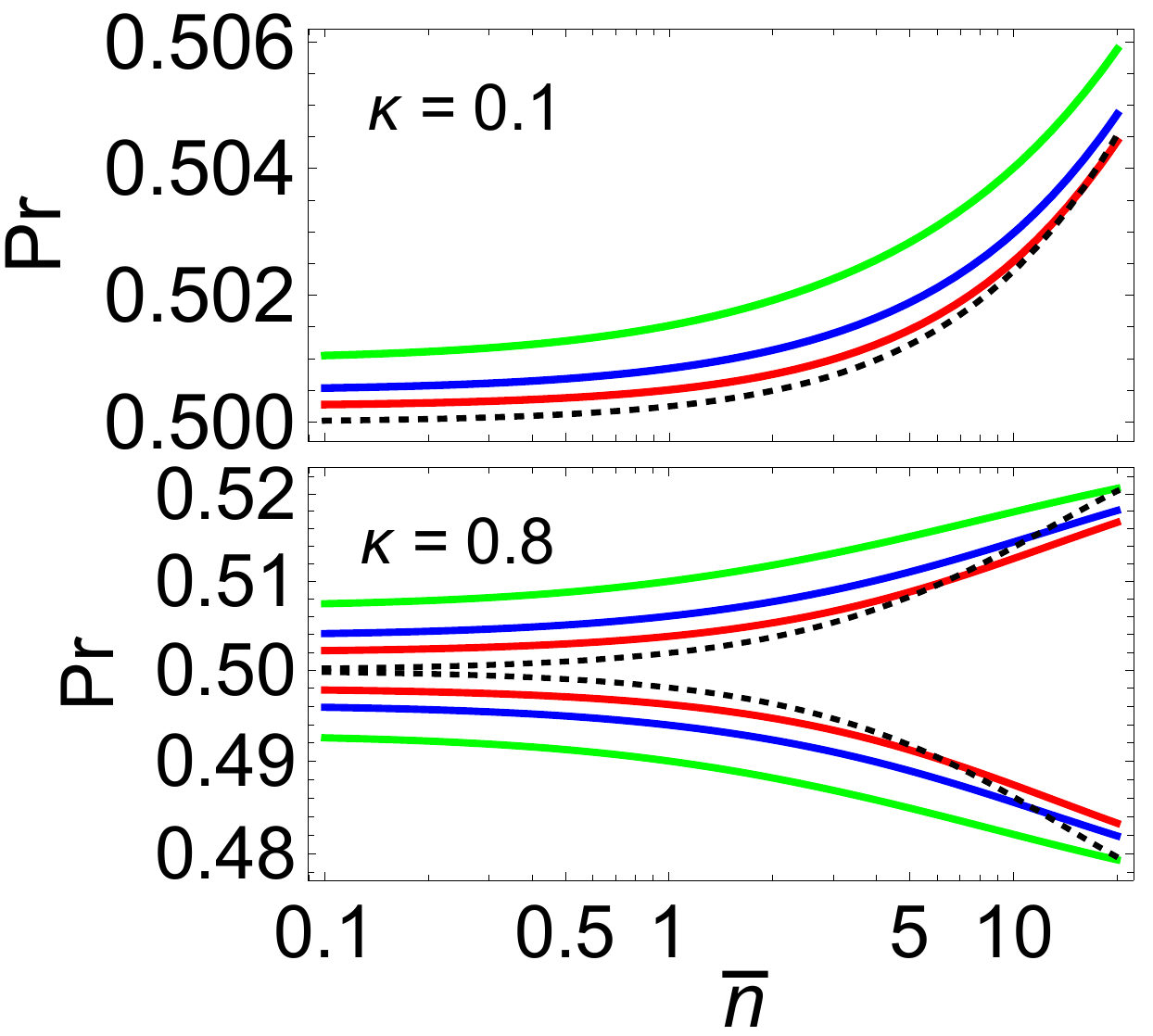}
	\caption{(Color online) Posterior probability $\Pr_1(H_1|1)$ after a click has occurred at a single receiving detector, calculated from the results from Fig.~\ref{fig:MultiplexedDetectors}, with assumed prior probabilities of $\Pr(H_0)=\Pr(H_1)=1/2$. Black dotted line is the performance of coherent state. From bottom to top, solid  lines above $\Pr=0.5$ correspond to $N=k$ heralded signal state: red (gray) for $\hat{\rho}_{1,(1,1)}$, blue (dark gray) for $\hat{\rho}_{1,(2,2)}$ and green (light gray) for $\hat{\rho}_{1,(4,4)}$; they follow a reverse order below $\Pr=0.5$. Object present probability is therefore correctly inferred because the thermal background state is less likely to cause a click compared to thermal plus signal.}
	\label{fig:PostMulticlickIdler}
\end{figure}

The advantage of quantum illumination compared with coherent state illumination eventually diminishes as $\nbar$ increases, up to a point where the coherent state outperforms multi-click heralded states. In the bottom plots of Fig.~\ref{fig:MultiplexedDetectors} and $\ref{fig:PostMulticlickIdler}$, there are crossover points between coherent state vs.~all quantum states, indicating that for mean photon numbers in excess of these crossover points using a coherent state for illumination is more likely to cause a signal detector click than multi-click heralded states. Such crossover points become more apparent for lower signal loss scenarios given $\kappa = 0.8$: heralded states generated by multiple simultaneous heralding clicks will show crossover points at higher $\nbar$, which can be explained by comparing the photon number distribution of the coherent state to those for multi-click heralded states. For high $\nbar$, the coherent state and its Poissonian photon number distribution continues to peak around the mean, whereas heralding click detection does not change the thermal distribution of TMSV much, meaning that after the target interaction process, the returning state $\hat{\rho}_1$ that contains a larger fraction of high photon numbers is more likely to trigger a click at the signal detector. For low $\nbar$, there is no crossover because the coherent state still contains vacuum probability whereas an $k$-click heralding detection removes probabilities from vacuum to $k-1$ photons, resulting in a significant change from a thermal photon number distribution. This low-photon advantage of quantum illumination is reflected in the nonclassicality criterion of the Fano factor: similar crossover points can be seen in Fig.~\ref{fig:FanoF} because the photon number variance increases with mean photon number.

Fig.~\ref{fig:PostMulticlickIdler} shows the single detector click probability along with respective posterior probability of target presence given click data, as a function of $\nbar$ for both the coherent state and the multi-click heralded state. These probabilities are calculated via Bayes' Law with no prior knowledge of the presence of the target so that $\Pr(H_1) = \Pr(H_0) = 1/2$ -- a click at the detector therefore increases our estimate of the probability that the target is present, and decreases if it is absent. Again, for low mean photon number there is a persisting enhancement to estimation, demonstrating the quantum illumination advantage. Plots of the probability for target present (absent) follow a similar behaviour as those for signal detector click probability both at low and high loss regimes. We use such posterior probabilities in Sec.~\ref{sec:dettraj} to inform a sequential click measurement process to estimate the presence or absence of the target. 

\begin{figure}[!t]
	\centering
	\includegraphics[width=8.6cm]{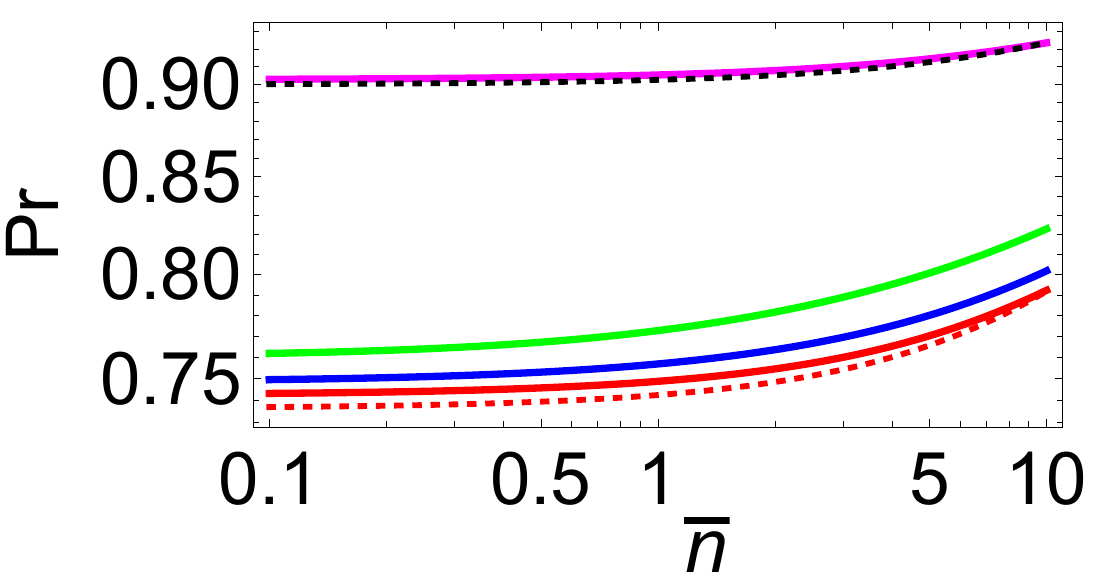}
	\caption{(Color online) Multi-click probabilities for a receiving detector multiplex, with detector efficiencies $\eta = \eta_S = 0.9$, background noise $\bar{n}_{B} = 10$ and object reflectivity $\kappa = 0.3$. The upper two curves are the single detector, single click probability $\Pr_1(1|H_1)$ for: black dotted - coherent state, and magenta (gray) for $\hat{\rho}_{1,(1,1)}$. The lower four curves are the double click probability $\Pr_2(2|H_1)$ for two detector multiplex: red dotted (coherent state), red (gray) for $\hat{\rho}_{1,(1,1)}$, blue (dark gray) for $\hat{\rho}_{1,(2,2)}$ and green (light gray) for $\hat{\rho}_{1,(4,4)}$.}
	\label{fig:MultReceiver}
\end{figure}

We also briefly show results where a click detector multiplex has been used as the receiving signal detector. In Fig.~\ref{fig:MultReceiver}, we show examples of double click probability at the signal detector as a function of mean photon number of the unconditioned signal. Overall, single click probabilities from a single detector are more likely to occur than double click probabilities. In order to see the effect of multi-click heralding and to separate its effect clearly from that of a coherent state return signal a slightly higher target reflectivity has been considered. For both sets of curves, the performance of quantum illumination with multiplexed heralding exceeds the bounds set by the coherent state illumination. However, the increase in click probability between quantum and coherent state illumination is lower for a single click because the difference in mean photon number between return states produced by a coherent state vs.~that produced by the state $\hat{\rho}_{1,1}$, is small.

\section{Click probability matching}
\label{sec:clickmatch}
In this section we outline a method of boosting the detectability of the multi-click heralded state in the quantum case without significantly compromising source discoverability, whilst enhancing click probability. From Fig.~\ref{fig:HeraldandAvg} we see that the coherent state produces the highest click probability for all mean photon numbers at a single click detector, compared to heralding probabilities for the TMSV. Consider a use case of the quantum illumination scheme in a quantum lidar, for which covertness of the signal is important, so mean photon numbers should be kept low. If we match the photon number of the TMSV to that of the coherent state, in order to compare quantum illumination with classical ``fairly", the TMSV will be more covert if a click detector is used by the target (or another party) to try to detect the use of the lidar. We may use this slight edge to our advantage by increasing the mean photon number of the TMSV above that of the coherent state so that the mean photon numbers are not matched but the click probabilities are. Then the {\it discoverabilities} of the TMSV and the classical coherent state are matched, at least up to a single click. The rest of this section outlines the consequences of this.

\begin{figure}[!t]
	\centering
	\includegraphics[width=8.6cm]{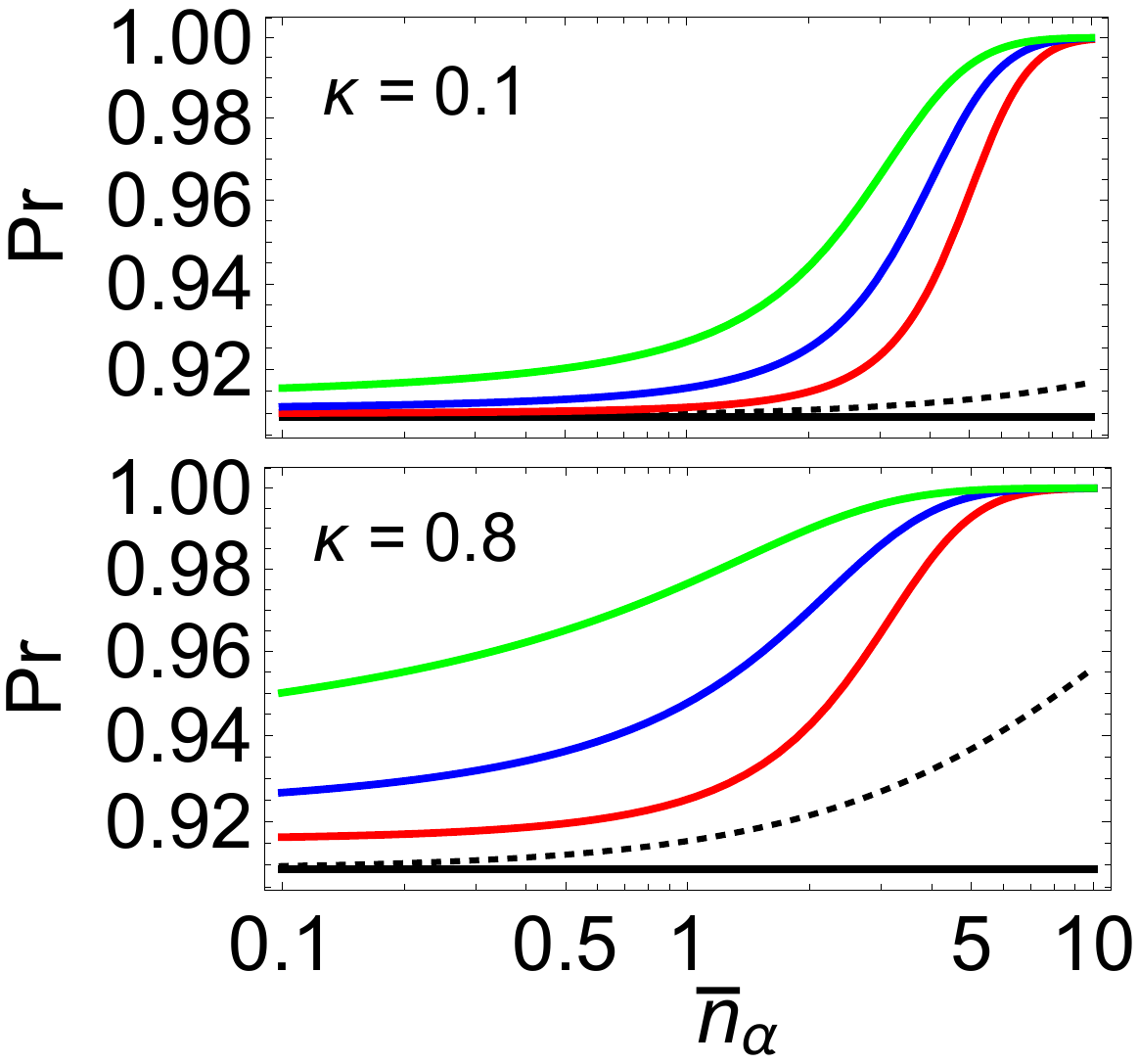}
	\caption{(Color online) Single receiving detector single click probabilities, using click probability matching, against mean photon number of a coherent state $\nbar_\alpha$. The parameters used to calculate the results of this figure are similar to that of Fig.~\ref{fig:MultiplexedDetectors}, however it is immediately evident that click probability matching enhances the performance of the quantum states compared to the coherent state (black dotted line). For both figures, the solid black line is the click probability $\Pr_1(1|H_0)$; the rest of the solid lines are: red (gray) for $\hat{\rho}_{1,(1, 1)}$; blue (dark gray) for $\hat{\rho}_{1,(2, 2)}$ and green (light gray) for $\hat{\rho}_{1,(4, 4)}$.}
	\label{fig:ClickProbMatch}
\end{figure}

\begin{figure}[!t]
	\centering
	\includegraphics[width=8.6cm]{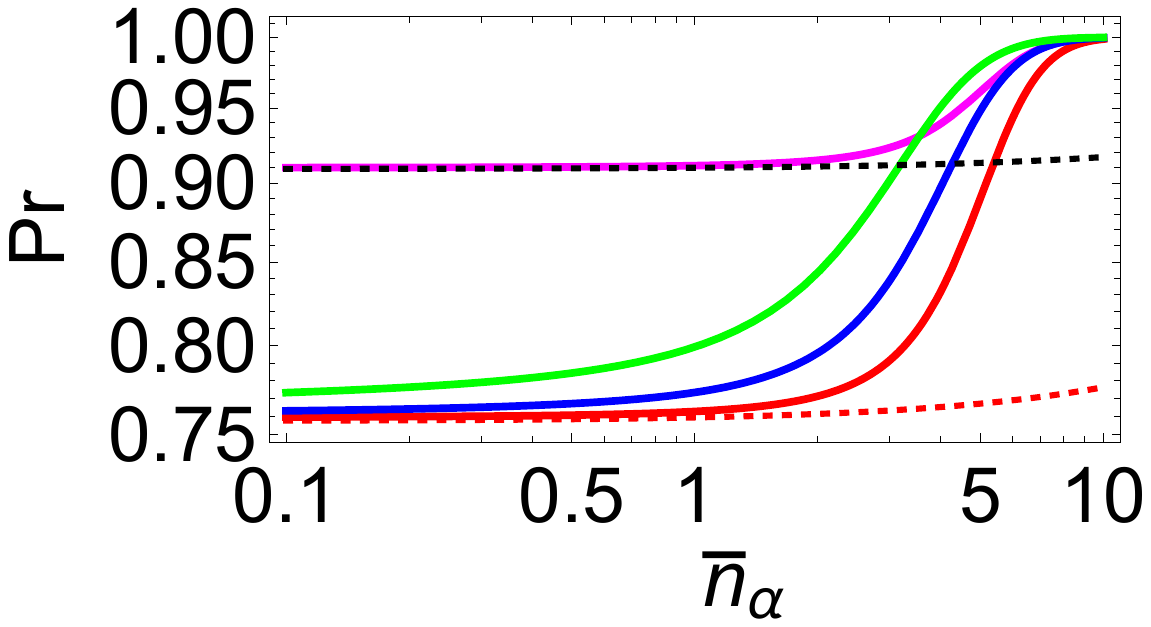}
	\caption{(Color online) Multi-click probabilities for a receiving detector multiplex (similar to Fig.~\ref{fig:MultReceiver}) showing click enhancement obtained by click probability matching, against mean photon number of a coherent state $\nbar_\alpha$ and object reflectivity $\kappa = 0.3$. The upper two curves are the single detector, single click probability $\Pr_1(1|H_1)$ for: black dotted - coherent state, and magenta (gray) for the state $\hat{\rho}_{1,(1,1)}$. The lower four curves from bottom to top are the double click probability for a two-detector multiplex, that is $\Pr_2(2|H_1)$, for: red dotted (coherent state), red (gray) for $\hat{\rho}_{1,(1,1)}$, blue (dark gray) for $\hat{\rho}_{1,(2,2)}$ and green (light gray) for $\hat{\rho}_{1,(4,4)}$.}
	\label{fig:MultiClickMatch}
\end{figure}

If one mode of the TMSV is observed independently of the other, it appears as a thermal state with $\nbar_S = \sinh^2 r$. Let us assume the existence of an eavesdropping third party, armed only with a single click detector with quantum efficiency $\eta_E$ that wishes to intercept the signal before it interacts with the target. The eavesdropping detector click probability is
\begin{equation}
    \tr(\hat{\Pi}_{1,1}\hat{\rho}_S) = \frac{\eta_E\nbar_S}{1+\eta_E\nbar_S},
\end{equation}
which follows easily from Eq.~\eqref{Eq:h0click}. This is lower than the single click probability produced by a coherent state for the same mean photon number,
\begin{equation}
    \tr(\hat{\Pi}_{1,1}\hat{\rho}_\alpha) = 1 - e^{-\eta_E\nbar_\alpha},
\end{equation}
if we were to set $ \nbar_S = \nbar_\alpha = |\alpha|^2$ as the coherent state mean photon number. Instead we keep the mean photon numbers different, but equate the above click probabilities
\begin{equation}
    \nbar_S = \eta_E^{-1}(e^{\eta_E\nbar_\alpha} - 1),
\end{equation}
which allows quantum illumination at higher intensity because the single eavesdropping detector cannot accurately distinguish whether the incoming state is coherent or thermal with a single click detector. The advantage obtained by doing this is not quantum, but it does offset the advantage that the coherent state has for for high mean signal photon numbers and high reflectivity, shown by both top and bottom plots of Figs.~\ref{fig:ClickProbMatch}. We show results similar to Figs.~\ref{fig:MultiplexedDetectors} and \ref{fig:MultReceiver} except with click probability matching. Both examples show a marked increase in receiver click probabilities, and in the bottom panel of Fig.~\ref{fig:ClickProbMatch} that for quantum illumination exceeds that for coherent state illumination significantly for mean photon numbers greater than one. Click probabilities for coherent state signals no longer exceed those for multi-click heralded signals. For the click multiplex receiving detector results in Fig.~\ref{fig:MultiClickMatch}, double click probabilities are also enhanced by click probability matching, such that they exceed single receiver click probabilities for signal mean photon number beyond two, although for single and double heralding clicks the double click probability (blue curve) approaches single heralding and single receiving click probability (magenta curve) at best. A general remark here is multiplexed receiver detection surpasses the single click receiver as long as there are more heralding clicks than receiver clicks in the high $\nbar$ regime, as evident from  Fig.~\ref{fig:MultiClickMatch}.

The click probability matching strategy gives quantum illumination an additional advantage when an eavesdropper is searching for a probe signal with a single click detector. If we know that the eavesdropper is doing this we are not only allowed to increase the energy of the TMSV, we are forced to do so to maximise our covertness. This enhances click probabilities at the receiver detector and gives quantum illumination an advantage over over classical for all values of $\nbar$. The increase in energy does not reveal to the eavesdropper whether we are using classical or quantum illumination because the single click detector cannot reliably distinguish the photon statistics of a coherent state from a thermal state with mean photon number tailored to match the single detector click probability of the coherent state. They will both cause the detector to click with the same probability. One could argue that the click matched state is in principle more distinguishable from the coherent state than the mean photon number matched. A simple examination of the photon probability distributions would support this. This extra distinguishability could be revealed if the eavesdropper has a detection system that is responsive to $g^{(2)}$ or a higher order coherence. However, such a detection system requires (at least) two simultaneous detections to occur reasonably often to reveal the increased mean photon number. This is highly unlikely in a realistic physical situation.

 
\section{Detection Trajectories}
\label{sec:dettraj}

In this section we simulate a simple quantum or classical illumination detection process, using the results of single shot click detections through a Monte-Carlo process that outputs a detection trajectory generated by sequentially updating an estimated probability of target object presence (in turn its absence as well). Click results at both the heralding detector and the receiving detector are used to update the probabilities $\pr(H_0)$ and $\pr(H_1)$ through calculation of posterior probabilities based on simulated click results, such that for each subsequent shot, we have the following updated estimation of target presence and absence
\begin{subequations}
    \begin{gather}
        \pr^{(M + 1)}\left(H_1\right) = \pr_{N_S}^{(M)}\left(H_1| k_S\right),\\
        \pr^{(M + 1)}\left(H_0\right) = \pr_{N_S}^{(M)}\left(H_0| k_S\right),
    \end{gather}
\end{subequations}
where $M$-th shot estimation is denoted by the superscript $\pr^{(M)}$, thus $\pr^{(M+1)}$ is the updated estimation. Initially, for $M = 0$, calculation of the posterior probability $\Pr^{(1)}$ begins by assumption of equal \emph{a priori} probabilities: $\Pr^{(0)}(H_0) = \Pr^{(0)}(H_1) = 1/2$. Such a process may be repeated until eventual convergence, or past a predefined threshold.

\begin{figure}[t!]
    \centering
    \includegraphics[width=8.6cm]{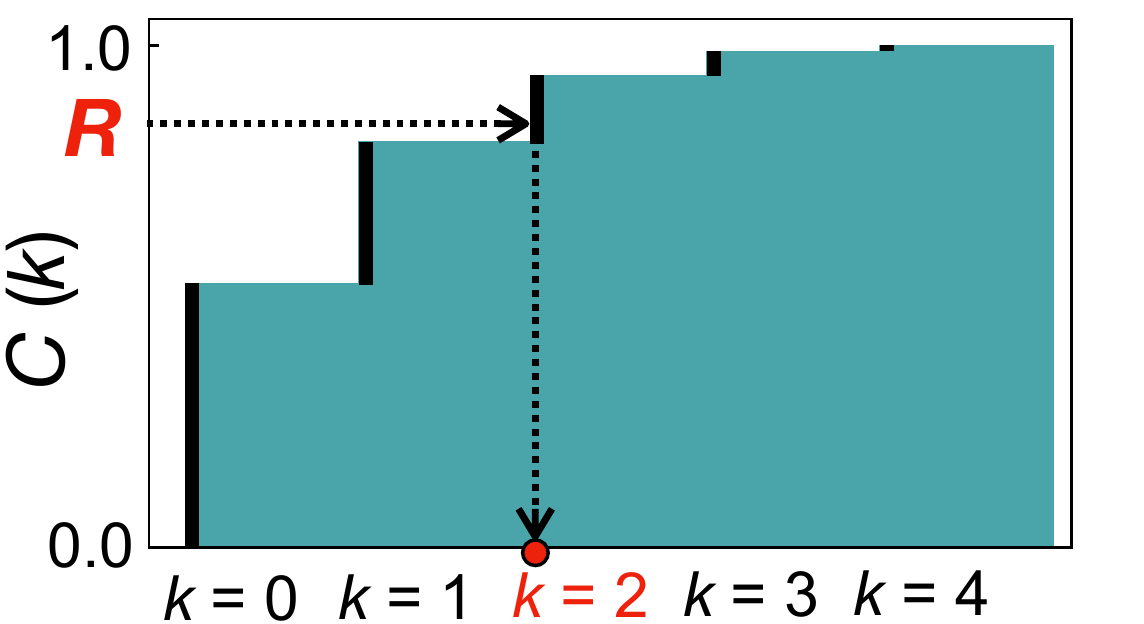}
    \caption{Diagram showing the distribution $C(k)$ and how clicks are selected by using a pseudorandom number $R$. This example shows cumulative heralding click probabilities from a four detector multiplex with $\nbar = 1$ and $\eta = 0.9$. Solid black edges represent $\Pr_{N,k}$. As $R = 0.82$, it then simulates $k_M = 2$ clicks because it falls within the interval covered by $\Pr_{N, 2}$.}
    \label{fig:clickselection}
\end{figure}

During each shot, probabilistic click results for the detector, $k_M$, are determined by inverse transform sampling where
\begin{equation}
    \label{Eq:selectionrules}
    k_M = C^{-1}(R),
\end{equation}
with $R$ being a pseudorandom number in the interval $(0,1)$. The function $C(k)$ is the discrete cumulative distribution of click probabilities for one detector
\begin{equation}
    C(k) = \sum_{k'=0}^k \tr(\hat{\Pi}_{N, k'}\hat{\rho}),
\end{equation}
for a multiplex of $N$ detectors measuring state $\hat{\rho}$. Rather than computing $C^{-1}(R)$ directly, the equality \eqref{Eq:selectionrules} can be achieved numerically by checking which probability interval $R$ falls between, and then assigning the corresponding $k_M$ (see Fig.~\ref{fig:clickselection}). Different signal states will evidently provide a different set of click probabilities, hence this will affect the random click sampling. The multi-click heralded states will be generated using this method: setting the click probability to the heralding probability in Eq.~\eqref{Eq:heraldprob} will allow a potentially different probe state $\hat{\rho}_{N, k = k_M}$ to be heralded for target detection as a more realistic test, because given TMSV of a fixed mean photon number, the heralding probability decreases for higher number of detectors in the multiplex hence it would be unlikely to generate states $\hat{\rho}_{N, k = N}$ for each shot.

To simulate detection results, the clicks of the signal detector measuring the background state, denoted as $k_{M0}$, are determined by sampling from the distribution
\begin{equation}
    C(k) = \sum_{k'=0}^k \tr(\hat{\Pi}_{N, k'}\hat{\rho}_0),
\end{equation}
which is the cumulative click distribution given the background state $\hat{\rho}_0$. Hence, the ``target absent" probability is always updated via
\begin{equation}
    \pr^{(M+1)}(H_0) = \pr^{(M)}_{N_S}\left(H_0|k_S = k_{M0}\right).
\end{equation}
To simulate detection results from illumination, given a static, present target, the clicks of the signal detector measuring the return state, denoted as $k_{M1}$, are determined by sampling
\begin{equation}
    C(k) = \sum_{k'=0}^k \tr(\hat{\Pi}_{N, k'}\hat{\rho}_1),
\end{equation}
such that ``target present" probability is updated via
\begin{equation}
    \pr^{(M+1)}(H_1) = \pr^{(M)}_{N_S}\left(H_1|k_S = k_{M1}\right).
\end{equation}
If we model quantum illumination using multi-click heralded TMSV signal, then generating the outcomes that correspond to $\hat{\rho}_{1} = \hat{\rho}_{1, (N, k = k_M)}$ requires an additional application of Eq.~\eqref{Eq:selectionrules} prior to detection (using a separate random number) conditioned by $C(k)$ with heralding click probabilities in order to simulate an independent ``heralding" process. For classical illumination using the coherent state signal, the return state $\hat{\rho}_1$ must be calculated using $\hat{\rho}_S = \ketbra{\alpha}$.

\begin{figure}[t!]
    \centering
    \includegraphics[width = 8.6cm]{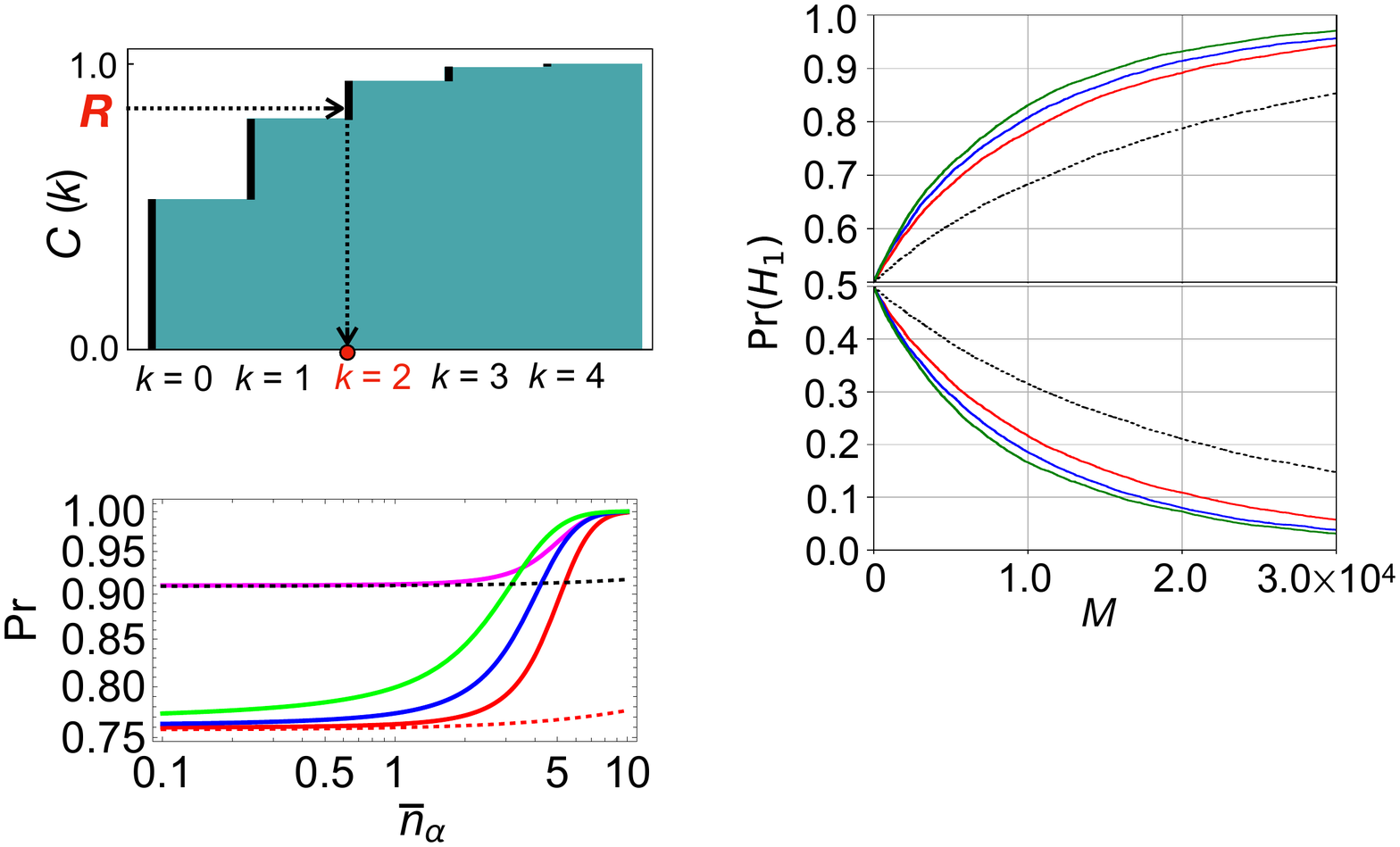}
    \caption{Trajectories of estimated object presence $\pr(H_1)$ from single receiving detector clicks, averaged over $3\times10^3$ observation trials. Each trial contains $M = 3\times10^4$ shots in order to show convergence of trajectories. Parameters: $\nbar = 1$, $\kappa = 0.1$ and $\nbar_B = 3$, heralding and receiving detectors have $\eta = \eta_S = 0.9$. Performance of coherent state $\ket{\alpha}$ is shown by dotted line. Solid lines from bottom to top: heralded state performances with $N$ detectors are: red (gray) for $N=1$, blue (dark gray) for $N=2$ and green (light gray) for $N=4$ (upper part of figure); they follow reverse order for lower part of the figure. Upper and lower parts of the figure demonstrates object presence and absence respectively -- quantum illumination correctly estimates state presence faster compared to coherent state illumination.}
    \label{fig:avg_trajectories}
\end{figure}

\begin{figure}[t!]
    \centering
    \includegraphics[width = 8.6cm]{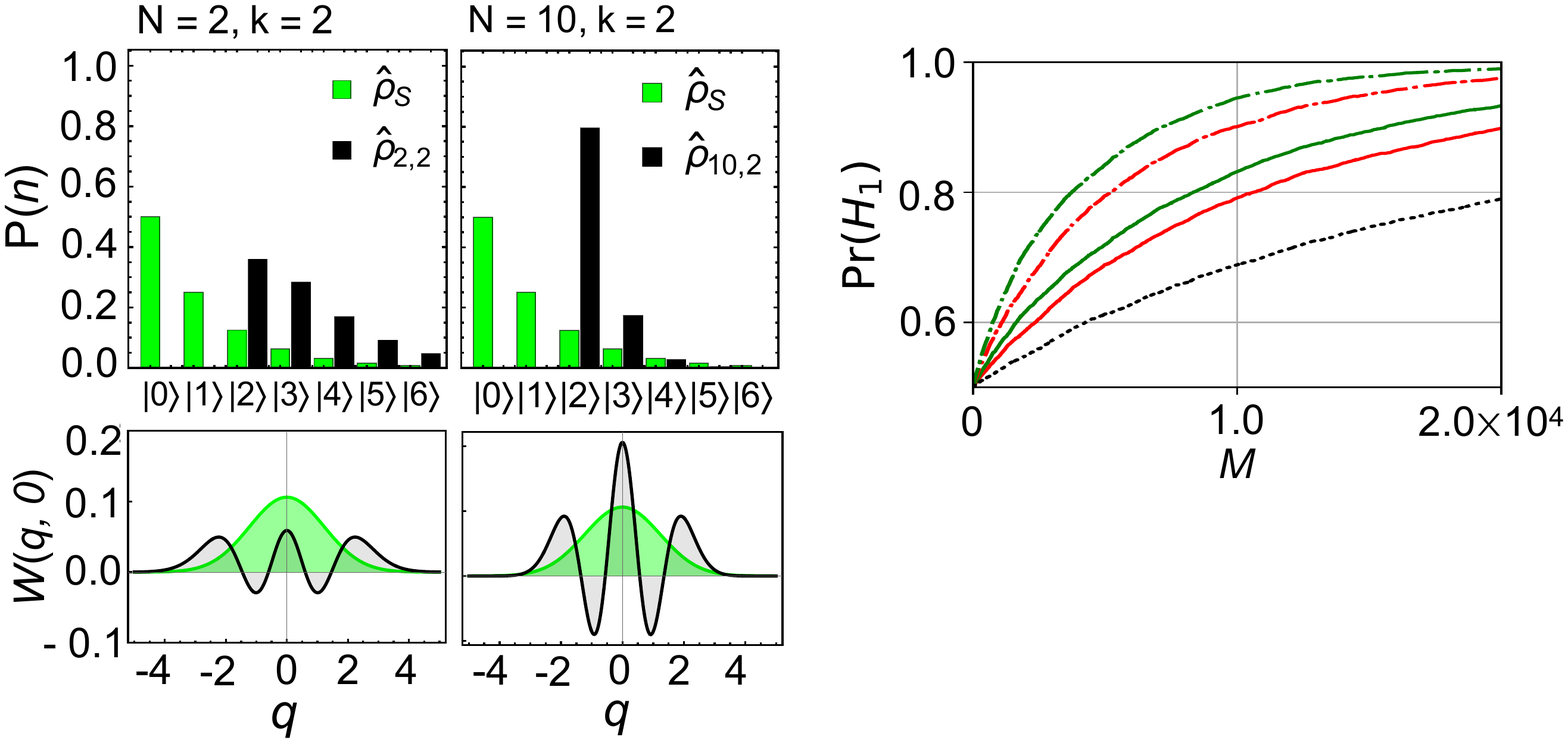}
    \caption{Trajectories using click probability matching averaged over $3\times10^3$ individual trajectories for a present object, each trial contains $M = 2\times10^4$ shots. Parameters: $\nbar = 1$, $\kappa = 0.1$ and $\nbar_B = 3$, heralding and receiving detectors have $\eta = \eta_S = \eta_E = 0.9$. Performance of coherent state $\ket{\alpha}$ is shown by dotted line. From bottom to top, solid lines are performances of heralded state with $N$ detectors: red (gray) for $N=1$, and green (light gray) for $N=4$; dash-dot lines are click probability matched average trajectories meaning that $\nbar=1.62$ for matched states and they follow the same order as the solid lines. Due to the increase in mean photon number from probability matching, matched multiplex click heralded TMSV will outperform the un-matched counterparts as click probability will be enhanced.}
    \label{fig:avg_trajectories_prmatch}
\end{figure}

This situation is shown in the upper half of Fig.~\ref{fig:avg_trajectories}, which gives an average of $6\times10^3$ trajectories for $3\times10^4$ consecutive shots of a quantum illumination experiment in which the mean photon numbers of the quantum beams are the same as the mean photon number of the classical beam. There are several things to note about this figure. Firstly the target object is present so any good illumination scheme must show a probability of target presence that tends to unity with increasing numbers of shots of the experiment. Secondly, the heralded quantum signals significantly outperform the classical in this respect. They reach higher probabilities of target presence after a significantly lower number of experimental shots. For example, a single idler detector heralded quantum source reaches a target present probability of 0.8 after 11166 shots, whereas the classical source takes 21386 shots, more than $~90 \%$ longer. Thirdly, the higher the number of detectors in our multiplex the better the quantum heralding. The curves in Fig.~\ref{fig:avg_trajectories_prmatch} seem to show little advantage in multiplexing but this is simply a perspective issue. The $N=1, 2, 4$ multiplex simulation curves reach an target present probability of 0.9 at 21045, 18092 and 15689 shots respectively. It takes $34\%$ longer to detect the target with a single detector than an $N = 4$ multiplexed idler.

For the simulation of target absence, the ``target present" probability is updated via the following
\begin{equation}
    \pr^{(M+1)}(H_1) = \pr^{(M)}_{N_S}\left(H_1| k_S = k_{M0}\right),
\end{equation}
hence the clicks generated from that of the background state, $k_{M0}$, are used to calculate the ``target present" probability, because the sent signal has been lost. The sequential process eventually accrues such a difference as the probability of target presence should decay with the number of shots, a situation shown in the averaged trajectory simulations of the bottom half of Fig.~\ref{fig:avg_trajectories}. The simulations also display the advantages of heralded quantum signals over classical and the increasing advantage of multiplexing, both of which are similar to the advantages displayed in the top half of the figure. Although there is no requirement for symmetry around the 0.5 value (the two plots relate to detection schemes operating in two different physical situations) the schemes appear to be equally as good at detecting a target when one is there as they are at ruling one out when it is not. 

There is one final point to note about the quantum schemes in Fig.~\ref{fig:avg_trajectories}. They provide a completely fair comparison with the coherent state, in that the heralded state is used no matter what the outcome of the idler detection. This is not typically the case in the analysis of heralded systems. When the idler detector does not fire, the state is sometimes discarded and another trial is run, typically at the pulse repetition rate of the pump laser producing the TMSV. This leads to a comparison between the coherent state and repeated successful heralding, which is unfair as the coherent state is unheralded and provides a successful pulse every time. 

We also show in Fig. \ref{fig:avg_trajectories_prmatch} the effect of click probability matching on our trajectories. We would expect this to decrease the number of shots taken to detect a present target (and the number of shots taken to rule out the presence of a target when one is not there). This is clearly shown in that the click-matched states reduce the detection times by about 50\%.

\section{Conclusions}
In this report we have provided a general framework for object detection using quantum states of light conditioned on multi-click heralding of the idler mode of a TMSV. Such a heralding and detection process is made possible by multiplexing so that the idler is split equally and falls onto an array of independent identical single APDs. These detectors are also fast and more robust against detector dead time constraints because there is a higher chance that other detectors may be active whilst some are ``dead" after photodetection.

Our results show that a small number of idler clicks from few detectors placed on the idler of TMSV heralds a nonclassical state, which has suppressed low photon number probabilities, a negative Wigner function and an increased mean photon number. The maximum possible increase in mean photon number for a set number of detectors is achieved when all detectors fire, though such events are rare compared to smaller numbers of clicks. The nonclassicality of the multi-click heralded state becomes less pronounced for higher values of unconditioned signal mean photon number, because the state becomes more thermal-like. 

The conditional boost to the signal number is the main effect exploitable for quantum illumination. The multi-click heralded TMSV proves useful for finding a target object hidden in high background thermal noise: the heralding allows use of a signal beam with an unconditioned mean photon number significantly lower than that of the background. The target presence is rendered more probable if the receiving click detector fires for a signal state with higher mean photon number as proportionately more signal will reflect from the target to the receiving detector. The conditional photon number enhancement provided by using such quantum states can not be obtained to the same extent from classical resources. In this work, we have mainly considered multiplexed detection for heralding and single click detection for receiving the reflected return signal. The advantages at low mean signal photon number are evident when the target reflectivity is low, or the thermal background noise is high. Our results demonstrate the increased advantage of using multi-click heralding states for target detection compared to single click heralding. The heralding advantage, however, persists only up to a particular mean photon number, higher than which it becomes better for target detection to use a coherent state. At higher signal intensity, the heralded state has close-to-thermal photon statistics and Poissonian states produces a more distinguishable signal in this regime. We can counteract the coherent state advantage at high photon numbers by matching the click probability of the average unheralded signal to that of the coherent state, which boosts the mean photon number of TMSV.

We have also modelled a sequential multi-shot hypothesis-testing procedure based on updating the prior probability of the target presence based on whether or not the signal and idler detectors fired. Results of sequential measurement were obtained using Monte Carlo simulation of detection statistics. Our update strategy not only includes using states when one or more of the idler detectors fires, but also when none of them do. This negates the need to wait for an idler firing event and allows for a fair comparison between quantum and classical state updates. The main result is that including the possibility of multi-click heralded states in the Monte-Carlo simulation reduces the target detection time significantly (i.e. number of measurement shots) compared with both single click heralding and classical illumination. Our modelling is done for relatively high object reflectivities, which are unrealistic in some real-world use cases such as lidars (where the fraction of signal photons reaching the detector might be $10^{-9}$). In such a scenario, to simulate enough trajectories to form a smooth average such that there are significant changes in probability would require billions of shots rather than thousands, which is easy to perform experimentally but impractical to simulate with available computational resources, and is likely to have similar-looking outcomes to Fig.~\ref{fig:avg_trajectories}, albeit on a different scale.

The main challenge in using quantum states for object detection is that for low reflectivity targets a long distance from the light source and detectors, the time taken to detect a useful number of reflected photons is long. The results shown here describe possible improvements that could lead to a usable future quantum version of lidar. However, this is still a significant challenge because the effective object reflectivity, governed by the inverse square law, will be tiny (for example around 10$^{-7}$ for an object 100m distant monitored by signal detectors of total effective area 0.01m$^2$). In order to change the probability that a target object is present there must be a some chance that a reflected photon is detected. Thus we require a number of shots of the experiment $M$ that satisfies $M/(\bar{n}\eta\kappa) \sim 1$ before we can change this probability in any significant way (so around 10 million in the example above, relatively quickly reachable at rapid optical pulse rates). This is true of both classical and quantum optical illumination schemes. The advantage quantum has is that for low $\bar{n}$ we gain significant knowledge from the idler detection results about which of these $M$ signal pulses are likely to contain significant numbers of photons and which are not. We can use this knowledge to bias our signal detection statistics to provide the quantum advantage shown here and detect the object in fewer pulses than classical or at a lower level of illumination.

The main result of this paper is that multi-click heralding increases the conditional intensity of the source, renders the detected counts more useful and therefore reduces the detection time. It can also negate some of the detector dead time effects at the idler detector so that heralded signal sources could be run more rapidly and at higher mean photon numbers. Our results suggest that with these improvements, at low mean photon numbers per shot, quantum illumination always outperforms classical coherent state illumination. The photon number per shot is currently limited technologically to be very low, so idler clicks are rare. The main problem is the low optical nonlinearity that is used to generate the TMSV states. This, when combined with a target reflectivity dominated by the inverse square law, will limit usage to tens of meters initially. If optical nonlinearity improves however, quantum lidar will become a possible, perhaps feasible, technique for detecting objects at a lower illumination level than is possible classically.

\acknowledgements
The authors would like to thank the the UK Engineering and Physical Sciences Research Council for funding via the UK National Quantum Technology Programme and the QuantIC Imaging Hub (Grant Number EP/T00097X/1), QinetiQ and the University of Strathclyde. They would also like to thank Jonathan D. Pritchard and Daniel Oi for useful discussions. 

\bibliography{library.bib}

\end{document}